\documentclass[review]{elsarticle}

\usepackage{lineno,hyperref,eurosym}

\journal{Acta Astronautica}

\usepackage{geometry}
\usepackage[english]{babel}
\usepackage[utf8]{inputenc}
\usepackage{newtxtext,newtxmath} %
\usepackage{enumitem} 
\usepackage[dvipsnames]{xcolor}
\usepackage{tikz}
\usetikzlibrary{positioning, matrix, decorations.pathreplacing, calligraphy, shapes.geometric}
\usetikzlibrary{arrows,automata}
\usepackage{tikz-3dplot}
\usepackage{cleveref}
\usepackage{tabularray}
\usepackage{nicefrac}
\usepackage{float}
\usepackage{makecell, amsmath, multirow}
\usepackage{tabularx}
\usepackage[table]{xcolor}
\definecolor{LightBlue1}{RGB}{180,210,240}
\definecolor{LightRed1}{RGB}{255,160,160}
\definecolor{SaturatedBlue}{RGB}{130,170,220}
\definecolor{SaturatedGreen}{RGB}{170,220,170}
\definecolor{SaturatedOrange}{RGB}{255,190,120}
\hypersetup{
    colorlinks,
    linkcolor={black!50!},
    citecolor={black},
    urlcolor={black}}
\usepackage{adjustbox}

\newcommand\setItemnumber[1]{\setcounter{enumi}{\numexpr#1-1\relax}}

\begin{document}

\begin{frontmatter}

\title{Low-Energy Round-Trip Trajectories to Near-Earth Objects\\ Using Low Thrust}

\author[label1]{Alessandro Beolchi}
\author[label2]{Mauro Pontani}
\author[label3]{Kathleen Howell}
\author[label1]{Chiara Pozzi}
\author[label4]{Sean Swei}
\author[label5]{Elena Fantino$^{*,}$}
\address[label1]{Department of Aerospace Engineering, Khalifa University of Science and Technology, P.O. Box 127788, Abu Dhabi, United Arab Emirates}
\address[label2]{Department of Mechanical and Aerospace Engineering, Sapienza Università di Roma, Rome, Italy}
\address[label3]{School of Aeronautics and Astronautics, Purdue University, West Lafayette, IN 47907, USA}
\address[label4]{Department of Aerospace Engineering and Khalifa University Space Technology and Innovation Lab, Khalifa University of Science and Technology, P.O. Box 127788, Abu Dhabi, UAE}
\address[label5]{Department of Aerospace Engineering and Polar Research Center, Khalifa University of Science and Technology, P.O. Box 127788, Abu Dhabi, United Arab Emirates, elena.fantino@ku.ac.ae}
\cortext[cor]{Corresponding author: elena.fantino@ku.ac.ae (Elena Fantino)}

\begin{abstract}
\footnote{Pre-print submitted to Acta Astronautica}
Near-Earth Objects (NEOs) are valuable targets for space exploration because of their accessibility, scientific significance, and valuable resources. Advances in trajectory design have uncovered efficient routes to these bodies, but systematic strategies for designing transfers between Earth and NEOs remain limited, especially when low thrust is employed. This work introduces a streamlined methodology that combines the Sun–Earth circular restricted three-body problem (CR3BP) and the Sun–spacecraft two-body problem (2BP) for designing low-energy round-trip trajectories to NEOs. The current implementation operates in the planar approximation, providing an efficient foundation for large-scale exploration of near-Earth space. Three-body manifold trajectories and transit orbits serve as natural pathways for Earth departure and return through the $L_1$ and $L_2$ libration points, while the heliocentric 2BP identifies spacecraft–NEO encounters through intersections of the respective elliptical orbits. This hybrid structure supports the generation of large collections of round-trip trajectories without relying on computationally intensive optimization procedures, enabling rapid preliminary mission design across broad NEO populations. To facilitate low-thrust trajectory design, rendezvous and takeoff maneuvers are initially modeled as impulsive. Then, instantaneous velocity changes are translated into low-thrust arcs to improve propellant efficiency. Round-trip transfers are obtained by combining compatible outbound and inbound branches under straightforward mission constraints. This modular structure is particularly well suited for designing complex mission architectures that cannot be systematically identified using conventional patched-conics methods. Applied to a representative NEO population, the method produces a large ensemble of round-trip trajectories with low launch and return energies, wide temporal flexibility, and competitive rendezvous and departure impulses when compared to existing 2BP solutions.
\end{abstract} 

\begin{keyword}
Near-Earth Objects \sep Trajectory design \sep Round-trip transfers \sep Low-energy trajectories \sep Low-thrust trajectories \sep Orbital maneuvers 
\end{keyword}

\end{frontmatter}

\section{Introduction}\label{INT} 

\indent Near-Earth Objects (NEOs) are Sun-orbiting asteroids (NEAs) and short-period comets with perihelion below $1.3$ au, whose orbits cross those of the inner planets \cite{CNEOSNEObasics}. Most NEOs' aphelia lie within a $5.2$ au sphere, whose radius is defined by the orbit of Jupiter. Despite significant efforts for the classification and detection of NEOs and the large body of research on the characterization of their dynamical evolution and origin, the interest in the design of efficient trajectories to these objects is rather recent. Missions to NEOs are valuable because these bodies are largely unchanged remnants of the early solar system, providing insights into planetary formation. In some cases, their Earth-approaching orbits lower propulsive requirements, making missions to accessible asteroids less costly than lunar trips on average \cite{korsmeyer2008into}. NEOs also carry valuable resources that are expected to support future space activities and economy. Lastly, crewed missions to NEOs can serve as important precursors to Mars exploration, offering crucial experience in interplanetary travel. Unmanned missions complement this by enabling flexible reconnaissance, sample return, and mining.\\ 
\indent The first spacecraft (S/C) to visit a NEO was NASA's NEAR Shoemaker, which landed on 433 Eros in 2001, revealing critical insights into the asteroid geometry and mass distribution \cite{dunham2002near}. This was followed by JAXA's Hayabusa mission, which returned samples from Itokawa in 2010 \cite{yoshikawa2021hayabusa}. CNSA's Chang'e 2 flew by Toutatis in 2012  \cite{gao2013near}, and JAXA's Hayabusa2 mission visited Ryugu between 2018 and 2020, returning samples to Earth in 2020  \cite{tsuda2020hayabusa2}. NASA's OSIRIS-REx collected material from Bennu in 2020, delivered it to Earth in 2023, and is now en route to Apophis \cite{lauretta2017osiris}. NASA’s DART mission, launched in 2021, successfully demonstrated asteroid deflection by impacting Dimorphos in 2022, and shortening its orbital period as a result \cite{rivkin2023planetary}. ESA’s Hera mission \cite{michel2022esa}, launched in 2024, characterized the outcome of the DART impact, deploying the CubeSats Milani \cite{ferrari2021preliminary} and Juventas \cite{goldberg2019juventas} in support of this investigation. The proposed Asteroid Redirect Mission, conceptualized in the 2010s but never launched, aimed to retrieve a boulder from a large NEA and bring it into lunar orbit, showcasing techniques for asteroid retrieval \cite{mazanek2015asteroid}. NASA’s NEA Scout, launched with Artemis I in 2022, is a smallsat equipped with a solar sail designed to rendezvous with a small Earth co-orbital asteroid \cite{lantoine2024trajectory}. These missions employed conventional trajectory design approaches such as direct Lambert arcs, patched-conics (PC) methods, gravity-assist maneuvers, and low-thrust arcs (see Table 1 of \cite{fantino2025direct}).

\textcolor{black}{A number of notable surveys, such as the Near-Earth Object Human Space Flight Accessible Targets Study (NHATS), have identified one- and two-way NEA trajectories for crewed and unmanned missions using the PC method, often in conjunction with full-precision ephemerides. Adamo et al.~\cite{adamo2010asteroid} performed one of the earliest systematic screenings, identifying several human-accessible NEAs through a three-stage filtering and round-trip evaluation framework. Strange et al.~\cite{strangenear} extended this effort to thousands of targets, mapping both rendezvous and sample-return opportunities under chemical and solar electric propulsion (SEP) assumptions. Subsequent studies explored hybrid propulsion architectures and programmatic considerations for crewed missions, including high-power SEP concepts~\cite{landau2011near}, technology- and crew-driven feasibility constraints~\cite{zimmer2011going}, and comparisons between chemical and hybrid systems~\cite{herman2014human}. The NHATS surveys~\cite{barbee2010comprehensive,barbee2013near} introduced a fully parametrized round-trip search over wide launch windows, cataloging millions of feasible opportunities under stringent mission-duration and re-entry constraints. A later study also examined the benefits of high-power SEP for reducing injected mass and expanding the set of accessible targets~\cite{de2015identifying}.}\\ 
\indent Beyond cataloging accessible NEAs, research has advanced trajectory design strategies for rendezvous, retrieval, and even transportation of asteroids to near-Earth space \cite{colasurdo2002missions,hasnain2012capturing,landau2013trajectories,llado2014capturing}. In this context, direct low-energy trajectories that leverage the natural dynamics of the Sun-Earth circular restricted three-body problem (CR3BP), in conjunction with low-thrust arcs, have received limited attention so far. Sánchez and Yárnoz \cite{sanchez2016asteroid} introduced a method that employs low thrust and stable invariant manifolds of Lyapunov and halo orbits to redirect objects from heliocentric orbits into Earth's vicinity. Jorba and Nicolás \cite{jorba2021using} examined the capture of NEAs using approximations of stable manifolds of hyperbolic tori linked to the $L_3$ point in the planar Earth-Moon-Sun bicircular problem. Mereta and Izzo~\cite{mereta2018target} investigated target-selection strategies for low-thrust missions beginning at the Sun–Earth $L_2$ point, comparing increasingly complex dynamical models to evaluate how the assumed trajectory representation influences the ranking of candidate NEAs. Machuca et al.~\cite{machuca2020high} examined the feasibility of autonomous 3U CubeSats to perform flybys of near-Earth asteroids after deployment from periodic orbits around the Sun–Earth $L_1$ or $L_2$ points. Flyby opportunities for years 2019–2025 were identified by intersecting low-energy CR3BP trajectories with NEO orbits. The study illustrates the potential of low-energy dynamics for low-cost small-satellite missions, but it remains inherently limited to CubeSat platforms and short-duration flybys. Topputo et al.~\cite{topputo2021envelop} developed a systematic screening methodology to identify near-Earth asteroids reachable by a low-thrust 12U CubeSat departing from the Sun–Earth $L_2$ point, while solving large sets of time- and fuel-optimal trajectory problems with a realistic propulsion model with variable thrust and specific impulse. Mascolo et al.~\cite{mascolo2021fast} developed a fast semi-analytical method based on Edelbaum-type approximations to estimate minimum-propellant transfers from Earth to NEOs with small eccentricity and inclination. The formulation is tailored to individual Earth-NEO transfer geometries and does not support systematic exploration across wide launch windows or large NEO populations.\\ 
\indent The above contributions illustrate the growing relevance of low-thrust trajectories for large-scale NEO exploration and low-energy techniques for NEO trajectory design, but also reveal a significant gap in the literature. Existing methods are often tailored to specific mission configurations, rely on optimization-intensive procedures, and in many cases limit the analysis to one-way transfers. However, a unified, scalable, and flexible strategy for systematically designing low-energy round-trip transfers involving multiple NEOs within a single framework, explicitly leveraging the dynamical structures of the Sun–Earth system, remains limited in the literature. Lastly, current approaches do not support the construction of more elaborate architectures, such as multiple NEO visits using libration point orbits as intermediate parking locations, or Earth escape/return through a libration point of choice.\\ 
\indent This study builds on two previous contributions that establish the foundation of its methodology. The first, presented at the 74$^{\rm th}$ International Astronautical Congress in Baku~\cite{canales2023design}, introduced a trajectory design framework to rendezvous with NEOs that combines the invariant structures of the Sun–Earth CR3BP with heliocentric 2BP dynamics and demonstrates its application to low-thrust transfers. \textcolor{black}{In that work, continuous thrust was modeled through sequences of small impulses connected by short Keplerian arcs (additional details on the method can be found in \cite{beolchiIAC75})}. Subsequently, Fantino et al.~\cite{fantino2025direct} formalized and expanded the technique into a systematic strategy for designing planar low-energy trajectories to rendezvous with NEOs on low-eccentricity, low-inclination orbits. In that formulation, manifold trajectories (MTs) and transit orbits (TOs) associated with periodic orbits around the collinear libration points $L_1$ and $L_2$ serve as dynamical gateways enabling low-energy one-way transfers from Earth to nearby NEOs.\\
\indent The present work advances these earlier developments in several key directions. Primarily, it extends the methodology to include an inbound branch, enabling the systematic design of complete Earth–NEO–Earth round-trip trajectories. The outbound and inbound segments of the transfer are computed independently and later combined, resulting in a modular, flexible, and streamlined framework that avoids burdensome optimization schemes. \textcolor{black}{Moreover, it introduces a thrust model that propagates the motion of the S/C in the heliocentric 2BP under continuous thrust, enabling a more efficient conversion of impulsive maneuvers into propelled segments}. The analysis is performed within the planar approximation, restricting applicability to low-inclination NEOs. This assumption is justified as an initial step toward low-propellant, low-energy transfers relevant for future cislunar and Sun-Earth libration point infrastructure and to small-payload missions where launch and return costs are critical. Exploration of targets on Earth-like orbits is prioritized because their trajectories smoothly intersect MTs and TOs, lowering the total impulse required by the mission. Nevertheless, the method is applicable to objects on more eccentric orbits. To this end, a round-trip mission to Apophis is investigated as an example.\\ 
\indent This manuscript is organized as follows. Section~\ref{DM} defines the dynamical model, introduces the planar patched-CR3BP/2BP method, and describes the target selection criteria. The trajectory design strategy is the subject of Sect.~\ref{TDS}. First, the mission design parameters are presented. Then, the procedure for computing outbound and inbound transfers is illustrated, with impulsive round-trip solutions emerging from the combination of outbound and inbound trajectories. Lastly, the method for replacing impulsive maneuvers with low-thrust arcs is described. The numerical results for both impulsive and low-thrust round-trip missions are presented and commented in Sect.~\ref{RD}, which also discusses the strengths and limitations of the technique as an alternative to PC methods, including an examination of mission architectures enabled by the proposed methodology as well as a mission to asteroid Apophis as an illustrative target outside the selected population. Concluding remarks and future directions are provided in Sect.~\ref{CR}. A preliminary version of this study was presented at the 76$^{\rm th}$ International Astronautical Congress~\cite{beolchiIAC76}. 

\section{Dynamical model}\label{DM} 

\indent In the Sun-Earth system, when the S/C is in the vicinity of Earth, its motion is governed by the simultaneous gravitational action of both Sun and Earth, and the Sun-Earth-S/C CR3BP is used. Conversely, when the S/C is far from Earth, the gravitational pull exerted by the planet becomes negligible compared to that of the Sun, and the Sun-S/C 2BP model can be adopted. The existence of two distinct dynamical models to describe the motion of the S/C requires the definition of an appropriate transition between the two. In this work, the selected boundary is a spherical surface centered at the Earth and referred to as transition sphere, which reduces to a transition circle (TC) in the planar approximation. Combining the CR3BP for accurate modeling near Earth with the 2BP for heliocentric transfers improves dynamical accuracy with respect to direct 2BP arcs, uncovering low-energy paths that would otherwise remain undetected. At the same time, the use of Keplerian orbits to approximate heliocentric motion enhances computational efficiency and streamlines the overall trajectory design strategy. 

\begin{table}
\centering
{\renewcommand{\arraystretch}{1.1}
\small
\caption{Physical parameters of the Sun-Earth system used in this work \cite{scaref}.}
\label{tab:sun_earth_parameters}
\begin{tabular}{l|l|l}
Symbol & Definition & Value \\ \hline
$\mu_S$          & \makecell[l]{Sun gravitational parameter} & $1.3271244 \cdot 10^{11} \, \nicefrac{\mathrm{km}^3}{\mathrm{s}^2}$ \\
$\mu_E$          & \makecell[l]{Earth gravitational parameter} & $3.9860044 \cdot 10^5 \, \nicefrac{\mathrm{km}^3}{\mathrm{s}^2}$ \\
$R_E$           & \makecell[l]{Earth equatorial radius} & $6378.1366 \, \mathrm{km}$ \\
$\mu$           & Sun-Earth mass ratio & $3.0034806 \cdot 10^{-6}$ \\
$d$             & Sun-Earth distance & $149597870.7 \, \mathrm{km} = 1 \, \mathrm{au}$
\end{tabular}
}
\end{table}

\begin{figure}
\centering
\includegraphics[width=0.65\linewidth]{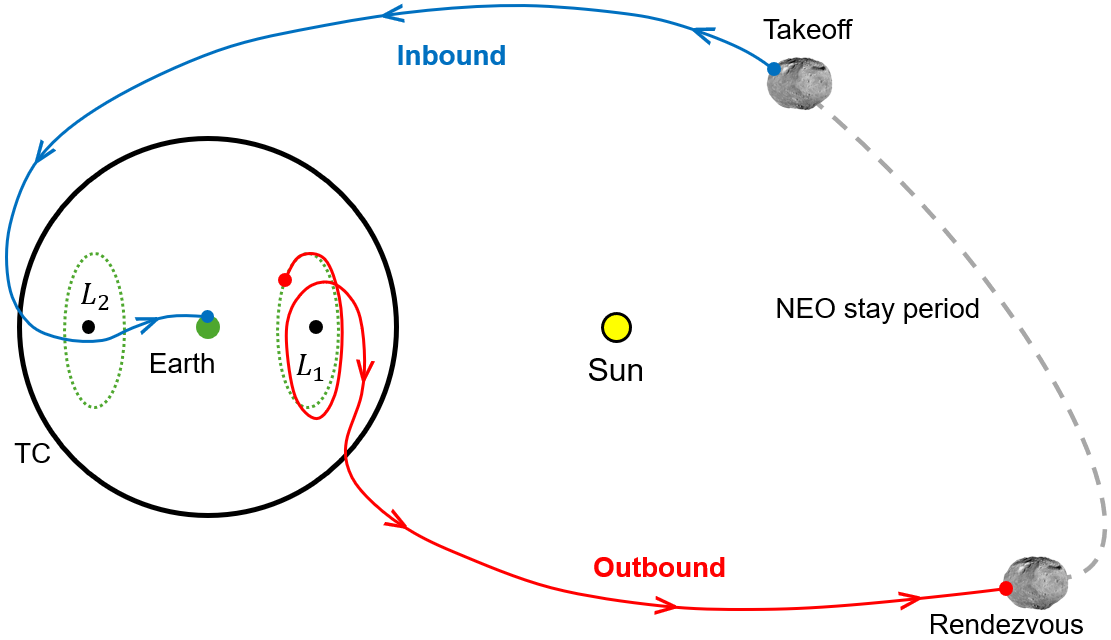}
\vspace*{-3.5mm}
\caption{Schematic representation of the trajectory design strategy showing the outbound and inbound phases connected through the NEO stay period.} 
\label{fig:strategy}
\end{figure}

\indent Using the sum \( \mu_S + \mu_E \) of the gravitational parameters of Sun and Earth and the distance \( d \) between the primary bodies as reference values, their orbital period equals \( 2\pi \). In the synodic barycentric reference frame, Sun and Earth are located at \( (\mu, 0, 0) \) and \( (\mu - 1, 0, 0) \), respectively, where \( \mu \) represents the mass ratio \( \mu_E / (\mu_S + \mu_E) \). The relevant physical parameters of the Sun-Earth system are listed in Table~\ref{tab:sun_earth_parameters}.\\
\indent The overall trajectory design strategy adopted in this work is outlined in Fig.~\ref{fig:strategy}. Specifically, the round trip mission architecture consists of an outbound segment departing from Earth, a stay period at the target NEO, and an inbound segment returning to Earth.

\indent Beyond the TC, the motion of the S/C follows 2BP dynamics, where trajectories emerging from the CR3BP transition into heliocentric Keplerian orbits. When returning to Earth, Sun-centered ellipses are transitioned to CR3BP trajectories as they cross the TC. As a consequence, the database includes two distinct sets of S/C orbits for outbound and inbound transfers. The fundamental building blocks of the database are families of planar Lyapunov orbits (PLOs) around the collinear equilibrium points $L_1$ and $L_2$. Each family includes $N_{LYAP} = 50$ orbits, uniformly distributed in Jacobi constant $C_J$, ranging from $C_J^{(\mathrm{min})} = 3.00056$ to $C_J^{(\mathrm{max})} = 3.00089$. The PLOs are used to generate MTs and TOs to reach the boundary of the TC. The remainder of this section draws from a previous contribution \cite{fantino2025direct}, summarizing only the main aspects of the low-energy three-body trajectories and dynamical model to ensure a clear and self-contained presentation.\\ 
\indent MTs are branches of unstable and stable hyperbolic invariant manifolds. After selecting a point on the PLO, the initial state of a MT is generated by applying a small perturbation in the direction of the unstable/stable eigenvector of the monodromy matrix of the departure orbit. MTs are obtained using standard techniques starting from $N_{MT} = 200$ points uniformly distributed in time along each PLO that constitutes the departure/arrival configuration for the outbound/inbound MT. Outbound/inbound MTs are globalized by propagating such initial states forward/backward in time until intersection with the TC occurs.\\ \indent TOs are trajectories contained within the manifold associated with a PLO and with the same energy level. TOs are constructed from a \( 70 \times 70 \) rectangular grid that encloses each PLO (in a way similar to that of \cite{fantino2025direct}). Each grid point located inside the PLO region provides the initial condition for a TO with the same Jacobi constant as the corresponding PLO. To simplify the analysis, the direction of the initial velocity is either parallel ($L_1$) or opposite ($L_2$) to the \( x \)-axis, ensuring that TOs move away from Earth along the corresponding unstable/stable manifold branch when propagated forward/backward in time, ultimately intersecting the TC. TOs are also propagated backward/forward in time toward Earth, and only those that match given departure/arrival conditions. 
    \begin{itemize}
        \item \textbf{Departure}: when propagated backward in time, the TO must intersect a 300-km altitude geocentric circular parking orbit;
        \item \textbf{Arrival}: when propagated forward in time, the TO must intersect the Earth surface.
    \end{itemize}
\noindent Because all MTs  are retained regardless of their approach to the Earth, unlike what occurs with TOs, the number of MTs is much greater than that of TOs.\\ 
\begin{figure}[t!]
	\centering
	\includegraphics[width=0.40\textwidth]{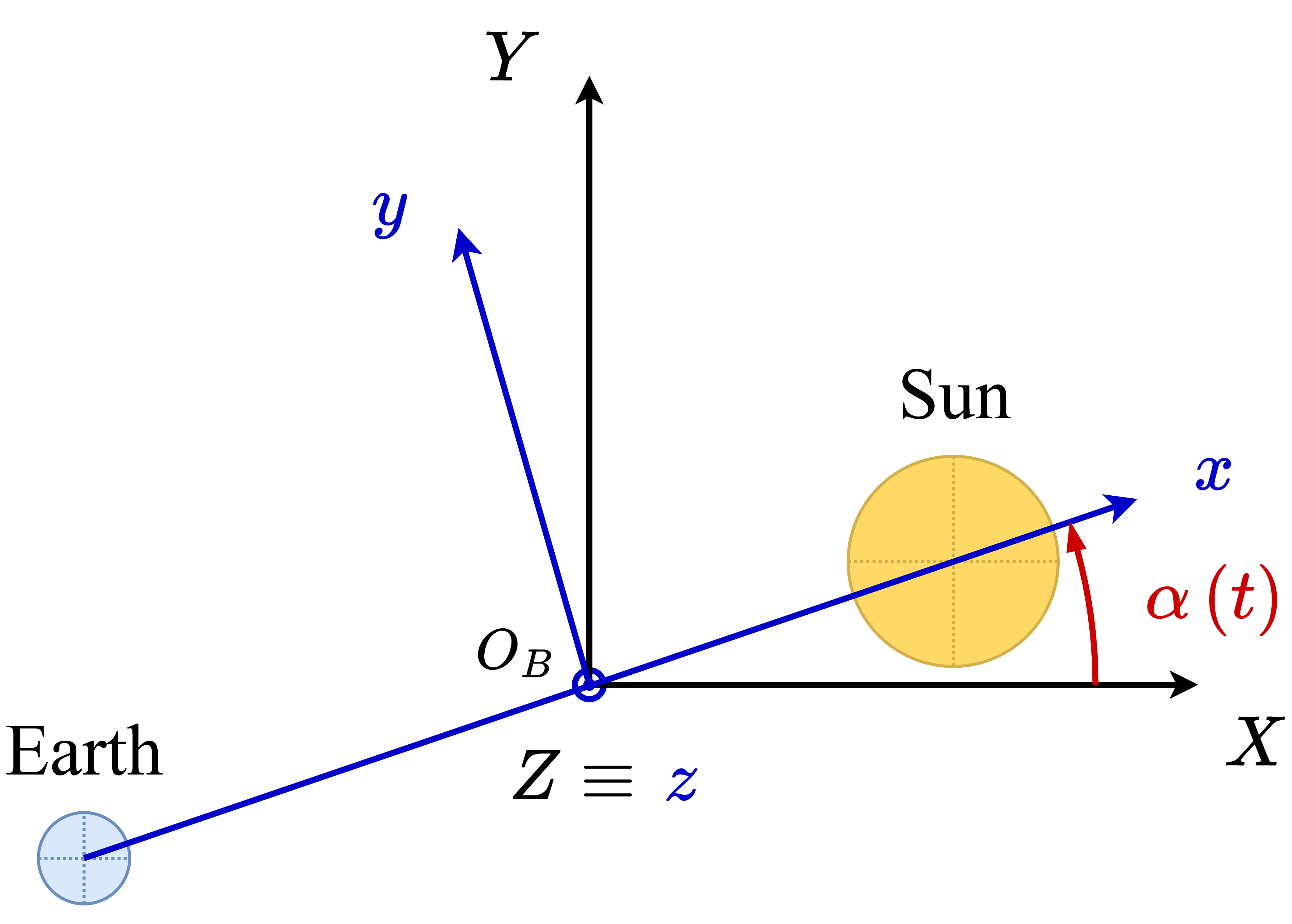}
	\vspace*{-3.5mm}
	\caption{Phase angle between the synodic ($x,y,z$) and inertial ($X,Y,Z$) barycentric reference frames.}
	\label{FIG:SI}
\end{figure}
\indent A TC with radius $R_{TC} = 2773940$ km, equivalent to approximately three times the radius of Laplace's terrestrial sphere of influence, is sufficiently large to contain all PLOs of interest and ensures that the manifold tubes cross the boundary of the TC transversely \cite{fantino2025direct}. MTs can be used to depart from (outbound) or arrive to (inbound) a PLO in the vicinity of an equilibrium point with negligible amounts of propellant, whereas TOs are direct trajectories from a geocentric circular parking orbit (outbound) or to Earth surface (inbound). The two types span the same energy levels, but while MTs initially stay close to the originating PLO, TOs pass through it and reach the TC directly, resulting in faster transfers. Therefore, TOs are much more attractive for short-duration transfers.\\ 
\indent At the TC, the state vectors of MTs and TOs are transformed to the heliocentric ecliptic J2000.0 inertial ($X_H,Y_H,Z_H$) reference frame. Due to the autonomous nature of the CR3BP, the same three-body trajectory is approximated by heliocentric orbits with the same semimajor axis and eccentricity, but different orientations, which changes with the phase $\theta_E$ of Earth, providing a useful degree of freedom for trajectory design. In particular, $\theta_E$ is a linear function of time,
\begin{equation}
    \theta_E \left( t \right) = \alpha \left( t \right) +\pi  = n_{SE} \left( t - t_{ref} \right) + \alpha_{ref} + \pi,
\end{equation}
\noindent where $\alpha$ is the phase angle between the synodic and inertial reference frame (see  Fig. \ref{FIG:SI}), $n_{SE}$ is the mean motion of the Sun-Earth system, $t_{ref}$ is a reference epoch, and $\alpha_{ref} = \alpha \left( t_{ref} \right) $. Consistently with previous work \cite{fantino2025direct}, the reference epoch is set to the Spring Equinox of 2023, $t_{ref} = 2460024.3917 \; \mathrm{JD}$,  corresponding to 2023-Mar-20 21:24:00 UTC, when the axes of the two reference frames are aligned (i.e., $\alpha_{ref} = 0$)\footnote{$\;$ The longitude of periapsis of the outbound and inbound orbits corresponding to MTs and TOs in the database is calculated using $t_{ref}$ as the departure and arrival epoch, respectively.}. The semimajor axis $a_{S/C}$, eccentricity $e_{S/C}$, and true anomaly $\theta_{S/C}$ are completely determined by the state vector at the TC in the synodic reference frame, while the longitude of periapsis $\varpi_{S/C}$ depends on the time of TC crossing, and varies at a constant rate equal to $n_{SE}$.

\section{Target selection}\label{TS}

\indent The first step in identifying low-energy Earth-NEO connections is to filter the available objects to retain only NEOs whose orbital elements are sufficiently close to those of Earth. This work relies on NASA's JPL Solar System Dynamics Small-Body Database \cite{SSDquery} to gather the most recent NEO data, which listed 38533 NEOs on June 23, 2025 \cite{CNEOSNEOstatistics}.

\begin{table*}
    \centering
    \caption{Candidate targets: primary designation and object index assigned in this work.} 
    \begin{adjustbox}{max width=\textwidth}
    {\renewcommand{\arraystretch}{1.1}
    \begin{tabular}{l r|l r|l r|l r|l r}
        Index & Designation & Index & Designation & Index & Designation & Index & Designation & Index & Designation \\
        \hline
         1 & 459872       & 17 & 2013 BS45   & 33 & 2019 GF1   & 49 & 2021 RZ3   & 65 & 2023 PZ     \\
         2 & 1991 VG      & 18 & 2013 GH66   & 34 & 2019 KJ2   & 50 & 2021 RG12  & 66 & 2023 RX1    \\
         3 & 2000 SG344   & 19 & 2013 RZ53   & 35 & 2019 PO1   & 51 & 2021 VH2   & 67 & 2023 RO16   \\
         4 & 2003 YN107   & 20 & 2014 DJ80   & 36 & 2020 CD3   & 52 & 2021 VX22  & 68 & 2023 XN13   \\
         5 & 2006 JY26    & 21 & 2014 QD364  & 37 & 2020 FA1   & 53 & 2022 BY39  & 69 & 2023 YO1    \\
         6 & 2006 QQ56    & 22 & 2014 WX202  & 38 & 2020 GE    & 54 & 2022 NX1   & 70 & 2024 BD4    \\
         7 & 2006 RH120   & 23 & 2015 JD3    & 39 & 2020 HF4   & 55 & 2022 OB5   & 71 & 2024 BY15   \\
         8 & 2008 KT      & 24 & 2015 XZ378  & 40 & 2020 HO5   & 56 & 2022 RS1   & 72 & 2024 DQ     \\
         9 & 2010 JW34    & 25 & 2016 RD34   & 41 & 2020 MU1   & 57 & 2022 RD2   & 73 & 2024 JQ1    \\
        10 & 2010 VQ98    & 26 & 2016 YR     & 42 & 2020 RB4   & 58 & 2022 RW3   & 74 & 2024 MS     \\
        11 & 2011 BL45    & 27 & 2017 BN93   & 43 & 2020 VN1   & 59 & 2023 BU7   & 75 & 2024 MM1    \\
        12 & 2011 MD      & 28 & 2017 FT102  & 44 & 2020 WY    & 60 & 2023 GQ1   & 76 & 2024 PD4    \\
        13 & 2011 UD21    & 29 & 2017 HU49   & 45 & 2021 AK5   & 61 & 2023 GT1   & 77 & 2024 PT5    \\
        14 & 2012 FC71    & 30 & 2018 PK21   & 46 & 2021 CZ4   & 62 & 2023 HL    & 78 & 2024 RA16   \\
        15 & 2012 LA      & 31 & 2018 PN22   & 47 & 2021 GM1   & 63 & 2023 HM4   & 79 & 2024 YL4    \\
        16 & 2012 TF79    & 32 & 2019 FV2    & 48 & 2021 LF6   & 64 & 2023 HG11  & 80 & 2025 DU7    \\
    \end{tabular}
    }
    \end{adjustbox}
    \label{tab:neo_candidates}
\end{table*}

\begin{figure*}
     \centering
     \includegraphics[width=0.85\textwidth]{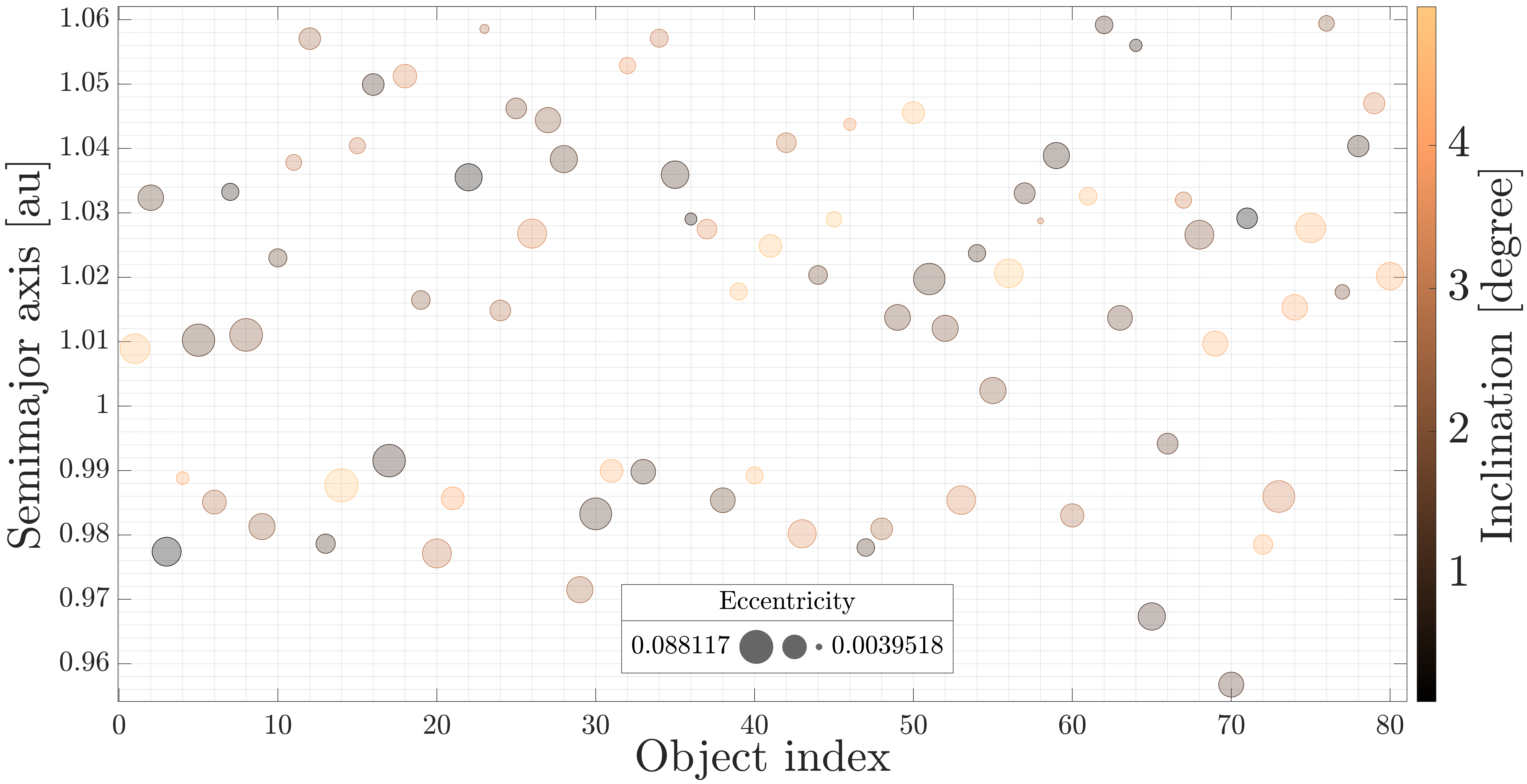}
     \vspace*{-3.5mm}
     \caption{Semimajor axis (y-axis), eccentricity (marker size), and inclination (color) of selected NEOs.} 
     \label{FigBubblechart}
 \end{figure*}

\begin{figure}
    \centering
    \includegraphics[width=0.40\textwidth]{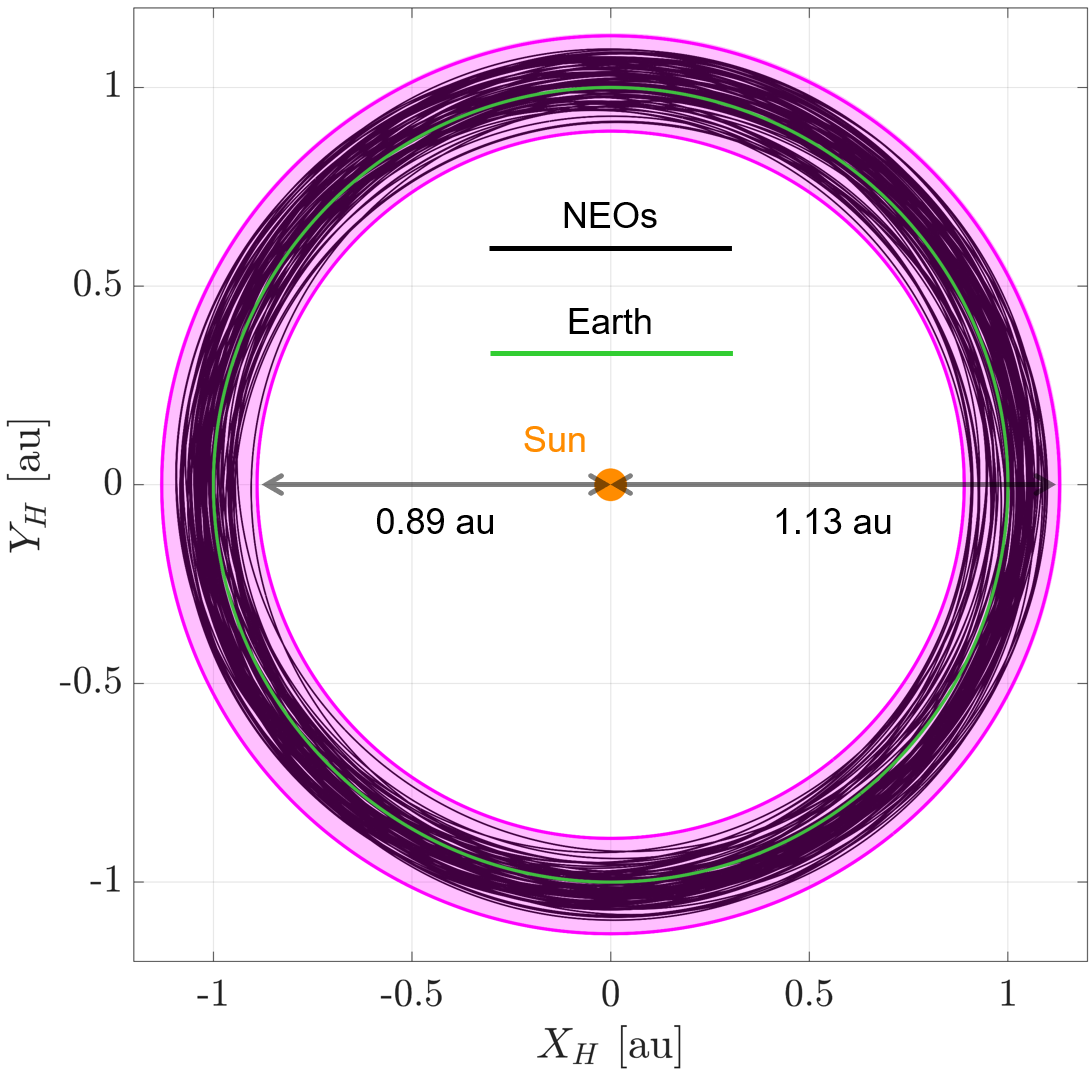}
    \vspace*{-3.5mm}
    \caption{Ecliptic view of the region of space accessed by MTs and TOs (delimited by purple lines) with the orbits of Earth and selected NEOs.}
    \label{FIG:ENEOsORBS}
\end{figure}
 
\indent In agreement with the fact that the S/C database orbits lie on the ecliptic plane and reach minimum perihelion distances of 0.89 au and maximum aphelion distances of 1.13 au, the population is restricted to objects with perihelion distance $q \geq 0.9$ au, aphelion distance $Q \leq 1.1$ au, and inclination $i \leq 5$ degree, yielding the set of 80 asteroids listed in Table \ref{tab:neo_candidates}. The adopted filtering criteria intentionally restrict the analysis to objects whose heliocentric orbits lie within the region accessed by the MTs and TOs (see Fig.~\ref{FIG:ENEOsORBS}). The S/C trajectories
exhibit modest eccentricity, remain in the proximity of the Earth orbit, 
and admit near-tangential intersections with the NEOs selected for this study. The inclination threshold for NEOs is introduced to ensure consistency between the planar approximation adopted here and the actual three-dimensional geometry of the target orbits. When three-dimensional transfers are addressed in a higher-fidelity model, the inclusion of low thrust can compensate for the neglected out-of-plane dynamics without incurring a significant $\Delta v$ penalty. This means that the mentioned filtering represents an intentional reduction of the object pool, aimed at identifying round-trip solutions compatible with limited propellant expenditure. Furthermore, the present work adopts the same selection criteria introduced in \cite{fantino2025direct} to ensure methodological consistency, while enabling a direct extension from one-way transfers to round-trip trajectories. \\
\indent The semimajor axes range from 0.9568 to 1.0594 au, the inclinations from 0.11 to 4.97 degree, and the eccentricities span the interval [0.0040, 0.0881]. Figure \ref{FigBubblechart} shows the semimajor axis $a_{NEO}$, eccentricity $e_{NEO}$, and inclination of the orbits of the target NEOs. All data are given in the heliocentric ecliptic J2000.0 reference frame, with osculation epoch $t_{osc} = 2460800.5$ JD, corresponding to 2025-May-05. Due to the modest inclination of the targets in the population, the orbital motion of the selected NEOs is approximated to planar, and the orientation of their 2BP orbits is given by the longitude of periapsis $\varpi_{NEO} = \Omega_{NEO} + \omega_{NEO}$. The ecliptic view provided by Fig. \ref{FIG:ENEOsORBS} shows that the orbits of Earth and selected NEOs are contained within the region of space accessed by MTs and TOs.\\ 
\indent Because most objects in the selected population are characterized by semimajor axes close to 1 au (see Fig. \ref{FigBubblechart}), the synodic period between Earth and any target asteroid is large, ranging from 12 to 276 years, with a 36-year average. As a result, the selection of departure and arrival time windows can have a significant effect on asteroid accessibility. In other terms, the choice of the departure epoch determines the number and set of NEOs reachable from Earth, while the choice of the arrival epoch affects the identification of NEOs compatible with a valid return segment. \textcolor{black}{The existence of outbound/inbound solutions in relation to the departure/arrival window represents a criticality of the target selection criteria, and will be examined in detail in Sect. \ref{RD}.} 

\section{Transfer design strategy}\label{TDS}

\begin{figure*}
    \centering
    \includegraphics[width=0.85\textwidth]{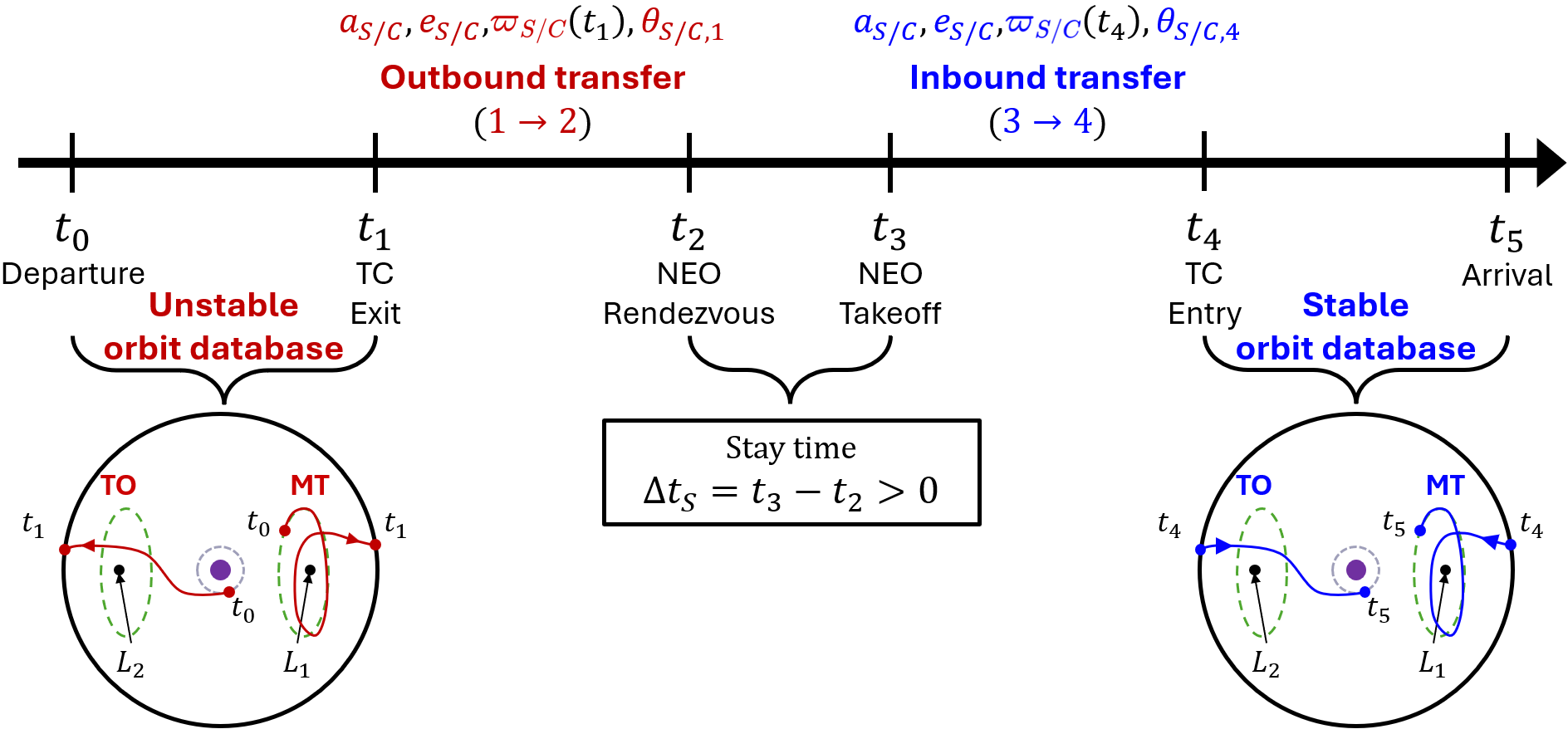}
    \vspace*{-3.5mm}
    \caption{Sequence of events of an Earth-NEO-Earth round-trip transfer designed with the patched-CR3BP/2BP method.}
    \label{FIG:STRAT}
\end{figure*}

\indent A round-trip transfer designed with the patched-CR3BP/2BP method is characterized by six events (see Fig. \ref{FIG:STRAT}). At $t_0$ the transfer begins. From $t_0$ to $t_1$ the S/C travels an unstable MT or TO. At $t_1$, it reaches and exits the TC. Between $t_1$ and $t_2$ the S/C follows an elliptical orbit around the Sun until it reaches the NEO. At $t_2$, rendezvous of the S/C with NEO occurs. In the interval $\left[t_2, \;t_3 \right]$ the S/C and NEO remain in rendezvous condition. At $t_3$, the S/C departs from NEO and is injected into another heliocentric ellipse leading to the TC. At $t_4$, the S/C enters the TC through a stable MT or TO, which is traveled until the final destination is reached at $t_5$. 

\indent The design of Earth-NEO-Earth round-trip transfers is based on four parameters, i.e.
\begin{itemize}
    \item departure epoch $t_0$;
    \item stay time at NEO $\Delta t_{S} = t_3 - t_2$;
    \item total mission duration (including $\Delta t_{S}$) $\Delta t = t_5 - t_0$;
    \item arrival epoch $t_5$.
\end{itemize} 
\noindent Some other relevant mission parameters are the total time spent inside and outside the TC, denoted by $\Delta t_{TC}$ and $\Delta t_{H}$, respectively. Note that $\Delta t = \Delta t_{TC} + \Delta t_{H} + \Delta t_{S}$.\\
\indent In the numerical implementation, $t_0$, $\Delta t_S$, and $\Delta t$ are used as independent design parameters. Lower and upper bounds are set on $t_0$, determining the departure window $\left[ t_0^{(\mathrm{min})}, \; t_0^{(\mathrm{max})} \right]$. Similarly, both stay time and total mission duration are constrained: $\Delta t_{S}^{(\mathrm{min})} \in \left[ \Delta t_{S}^{(\mathrm{min})}, \; \Delta t_{S}^{(\mathrm{max})} \right]$ and $\Delta t^{(\mathrm{min})} \in \left[ \Delta ^{(\mathrm{min})}, \; \Delta t^{(\mathrm{max})} \right]$. In the proposed methodology, outbound and inbound solutions are computed separately, and only at a later stage combined with an appropriate matching strategy. To obtain the inbound segment independently from the outbound segment, an arrival window $\left[ t_5^{(\mathrm{min})}, \; t_5^{(\mathrm{max})} \right]$ must be specified for the computation of inbound trajectories, just as a departure window is required for the calculation of outbound transfers. Note that $t_5^{(\mathrm{min})} = t_0^{(\mathrm{min})} + \Delta t^{(\mathrm{min})}$ and $t_5^{(\mathrm{max})} = t_0^{(\mathrm{max})} + \Delta t^{(\mathrm{max})}$.\\ 
\indent The selection of a suitable value for $\Delta t^{(\mathrm{min})}$ plays a critical role in the implementation of the solution strategy. Because $\Delta t^{(\mathrm{max})} \geq \Delta t^{(\mathrm{min})}$, it follows that $\Delta t_5 = t_5^{(\mathrm{max})} - t_5^{(\mathrm{min})} \geq \Delta t_0 = t_0^{(\mathrm{max})} - t_0^{(\mathrm{min})}$. However, the larger return window does not necessarily translate into higher computational cost for inbound trajectories, since the wider window is offset by the lower number of objects for which inbound transfers are investigated (i.e., the subset of NEOs accessed by outbound solutions). A small $\Delta t^{(\mathrm{min})}$ increases the number of inbound trajectories and potentially produces several infeasible outbound-inbound combinations with $t_3 - t_2 < \Delta t_S^{(\mathrm{min})}$. A large $\Delta t^{(\mathrm{min})}$ reduces the number of inbound segments and potentially overprunes the solution space by neglecting attractive short-duration transfers. \textcolor{black}{This means that, the choice of $\Delta t^{(\mathrm{min})}$ regulates the trade-off between computational efficiency and thoroughness in the exploration of the solution space.}

\subsection{One-way transfer}\label{OT} 

\indent Both outbound and inbound segments are one-way transfers. The outbound segment consists of the portion of the mission between TC exit and NEO rendezvous, covering the time frame from $t_1$ to $t_2$. The inbound transfer consists of the portion of the mission between departure from NEO and TC entry, covering the time frame from $t_3$ to $t_4$. The procedure for computing outbound transfers (see \cite{fantino2025direct}) is the following: 

\begin{figure}
     \centering
     \includegraphics[width=0.40\textwidth]{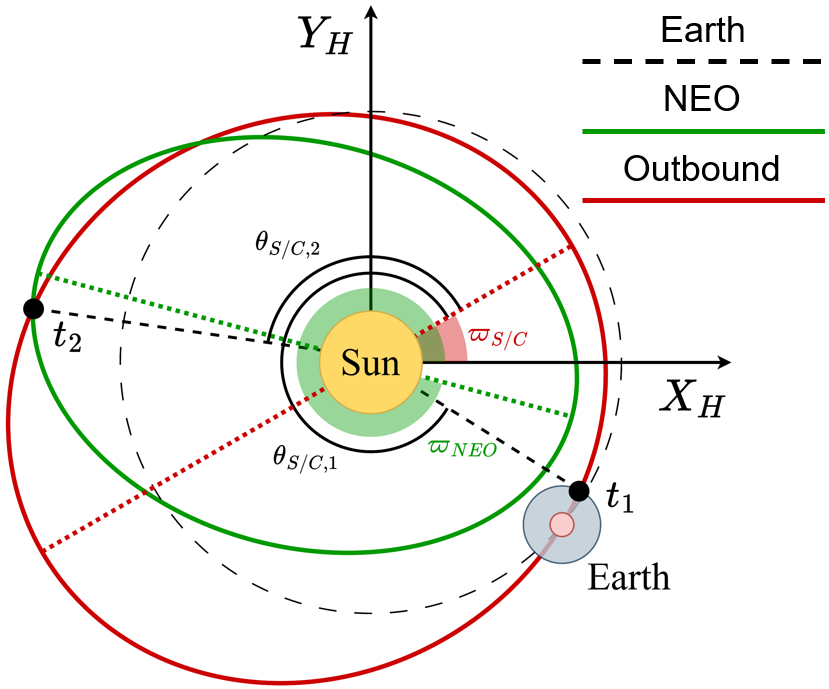}
     \vspace*{-3.5mm}
     \caption{Illustration of a planar one-way Earth-NEO (outbound) transfer.}
     \label{FigSchemeOWout}
\end{figure}

\begin{figure}
     \centering
     \includegraphics[width=0.40\textwidth]{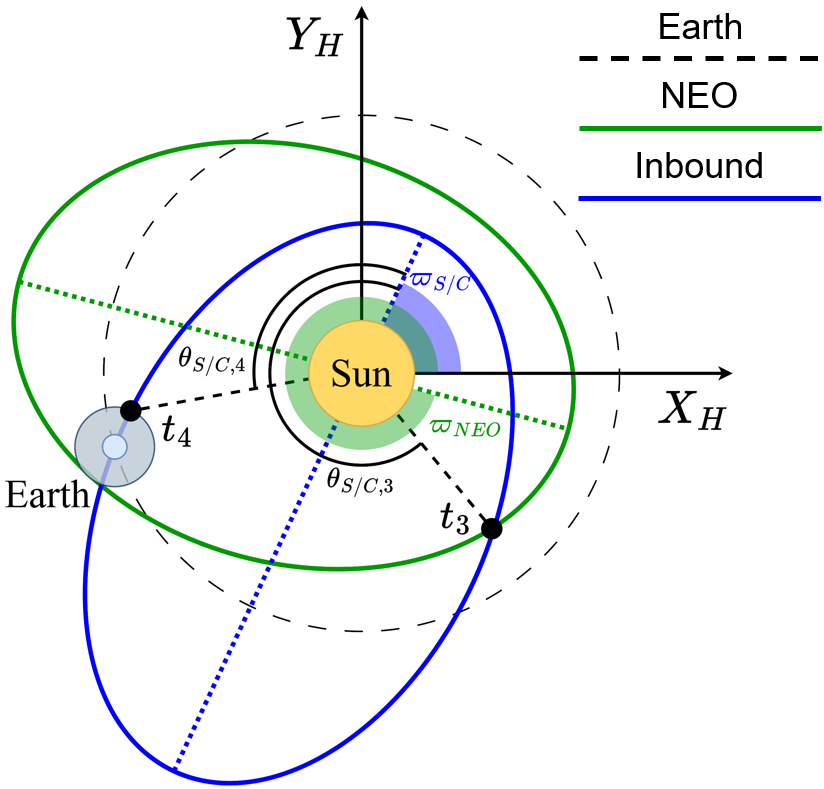}
     \vspace*{-3.5mm}
     \caption{Illustration of a planar one-way NEO-Earth (inbound) transfer.}
     \label{FigSchemeOWin}
\end{figure}

\begin{enumerate} 
    \item Given $\left\{ a_{NEO}, e_{NEO} \right\}$ and $\left\{ a_{S/C}, e_{S/C} \right\}$, and provided that at least one intersection between the two ellipses exists, the range of values of $\varpi_{S/C}$ for which intersections with the target orbit exist can be determined analytically (see \cite{wen1961study}). The range is sampled with constant step $\delta \varpi_{S/C}$;
    \item At the $i$-th step, $\varpi_{S/C}^{i)}$ is used to compute the intersection points between the elliptical orbits (if the intersection points are two and distinct, they are addressed individually). In particular, the true anomaly of the S/C at the intersection point $\theta_{S/C,2}$ and the rendezvous impulse $\Delta v_2$ are calculated;
    \item The passage of NEO through the intersection point determines the rendezvous epoch $t_2$;
    \item From $\left\{ a_{S/C}, e_{S/C}, \theta_{S/C,1},\theta_{S/C,2} \right\}$ (note that the true anomaly of the S/C at TC exit $\theta_{S/C,1}$ comes from the S/C orbit database), the time of flight along the outbound transfer $\Delta t_{12} > 0$ is calculated, and $t_1 = t_2 - \Delta t_{12}$;
    \item Once $t_1$ is known, the true longitude of the periapsis of the orbit of the S/C at TC exit $\varpi_{S/C}$ is computed. In general, $\varpi_{S/C} \neq \varpi_{S/C}^{i)}$, and $\Delta \varpi_{S/C} = \left| \varpi_{S/C} - \varpi_{S/C}^{i)} \right|$. A non-zero $\Delta \varpi_{S/C}$ translates into a non-zero rendezvous distance $\Delta r_2$ between NEO and S/C at $t_2$;
    \item The process is repeated $\forall \; \varpi_{S/C}^{i)}$, and only solutions with $\Delta v_2 < \Delta v_2^{(\mathrm{max})}$ and $\Delta \varpi_{S/C} < \Delta \varpi_{S/C}^{(\mathrm{max})}$ are retained. 
\end{enumerate}

\noindent The procedure for computing inbound transfers is symmetric. The process is repeated with the following adjustments, referred to the preceding points:

\begin{enumerate}
    \setItemnumber{2}
    \item $\theta_{S/C,3}$ and takeoff impulse $\Delta v_3$ (in place of $\theta_{S/C,2}$ and $\Delta v_2$);
    \setItemnumber{4}
    \item takeoff epoch $t_3$ (in place of $t_2$);
    \setItemnumber{5}
    \item $\left\{ a_{S/C}, e_{S/C}, \theta_{S/C,3},\theta_{S/C,4} \right\}$, TC entry $\theta_{S/C,4}$, time of flight along the inbound transfer $\Delta t_{34} > 0$ is calculated, and $t_4 = t_3 + \Delta t_{34}$;
    \setItemnumber{6}
    \item $t_4$ (in place of $t_1$), takeoff distance $\Delta r_3$ between NEO and S/C at $t_3$ (in place of $\Delta r_2$);
    \setItemnumber{7}
    \item $\Delta v_{3} < \Delta v_{3}^{(\mathrm{max})}$.
\end{enumerate}

\noindent Inbound trajectories are explored only for objects that admit at least one outbound solution. Sketches of outbound and inbound transfers are provided in Figs. \ref{FigSchemeOWout} and \ref{FigSchemeOWin}, respectively. 

\subsection{Round-trip transfer}

\begin{figure}
     \centering
     \includegraphics[width=0.40\textwidth]{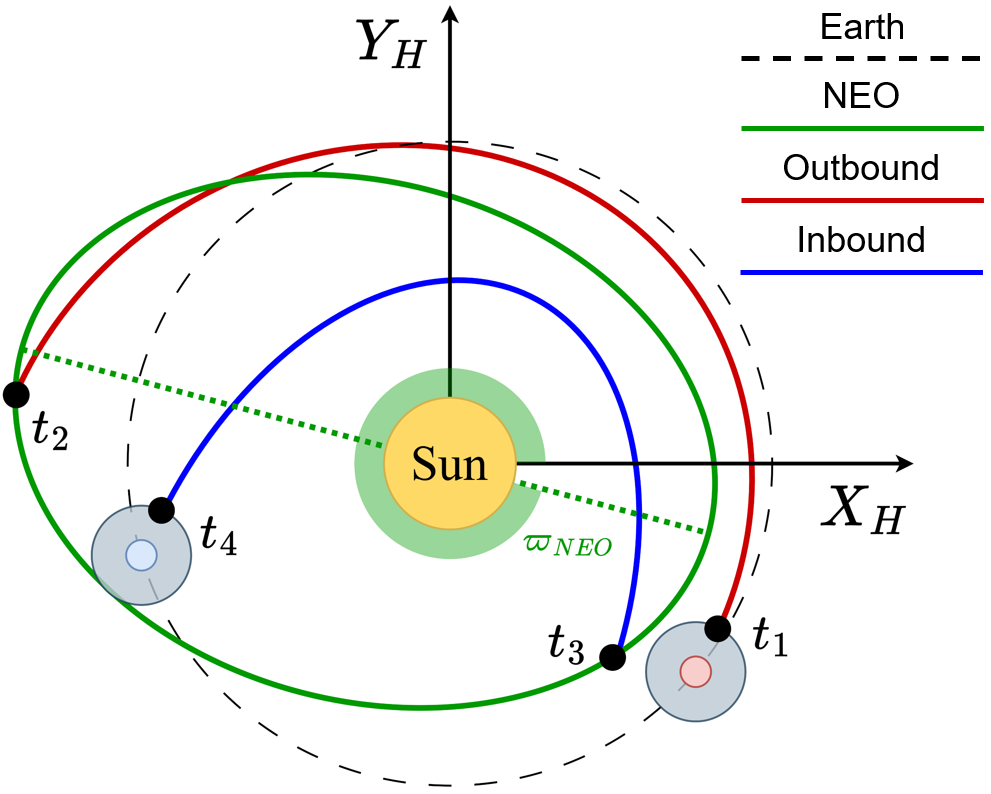}
     \vspace*{-3.5mm}
     \caption{Illustration of a planar round-trip Earth-NEO-Earth transfer.}
     \label{FigSchemeTW2D}
\end{figure}
\indent Round-trip transfers are obtained by combining outbound and inbound solutions to the same object. First, a subset of asteroids that can be reached with outbound transfers is found. Then, the objects for which inbound trajectories exist are identified. Each combination of outbound and inbound segments determines the stay time $\Delta t_S = t_3 - t_2$ and the total transfer time $\Delta t = t_5 - t_0$ for that specific round-trip transfer, which must satisfy mission-specific constraints for the solution to be considered feasible. A sketch of round-trip transfer is provided in Fig. \ref{FigSchemeTW2D}.\\
\indent A viable approach for eliminating the position errors at NEO rendezvous and takeoff introduced by the procedure based on intersection of ellipses consists in solving the Lambert problem between the state of the S/C at the TC and that of NEO at specified epochs. Preserving the event dates from the ellipse intersection strategy, two Lambert arcs are identified:
\begin{itemize}
    \item outbound leg, from TC exit (at epoch $t_1$) to NEO (at $t_2$), and
    \item inbound leg, from NEO (at $t_3$) to TC entry (at $t_4$).
\end{itemize}
In this case, four velocity changes occur at $t_i$ $(i=1,\ldots,4)$: $(\Delta v_1,\Delta v_2)$ and $(\Delta v_3,\Delta v_4)$, each pair associated  with the outbound leg and the inbound path, respectively. When the positional mismatch is small, both $\Delta v_1$ and $\Delta v_4$ remain negligible compared to $\Delta v_2$ and $\Delta v_3$, which determine the total velocity change of the round-trip transfer
\begin{equation}
    \Delta v = \Delta v_1 + \Delta v_2 + \Delta v_3 + \Delta v_4 \approx \Delta v_2 + \Delta v_3.
\end{equation}

\subsection{Finite-thrust Lambert solver}\label{ILT}

\indent To reduce propellant consumption, the instantaneous velocity changes associated with either an outbound or an inbound leg are replaced by low-thrust arcs. To do this, a dedicated conversion algorithm is designed and used on each leg. The final goal is to obtain two low-thrust arcs in place of instantaneous velocity changes, namely, (a) an initial thrust arc of duration $\Delta t_{ini}$, to leave the initial orbit, and (b) a final thrust arc of duration $\Delta t_{fin}$, to drive the spacecraft to the final specified position with the desired velocity. The algorithm at hand preserves the event dates (either $t_1$ and $t_2$ for the outbound leg or $t_3$ and $t_4$ for the inbound leg), as well as the endpoint states (position and velocity) corresponding to the impulsive transfer trajectory. 

\indent The problem consists of connecting an initial state $(\boldsymbol{r}_{ini},\boldsymbol{v}_{ini})$ at $t_{ini}$ ($=t_1$ or $t_3$) to a final state $(\boldsymbol{r}_{fin},\boldsymbol{v}_{fin})$ at $t_{fin}$ ($=t_2$ or $t_4$), using finite thrust. The thrust magnitude $T$ and the effective exhaust speed $c$ are  constant parameters that describe the propulsion system. If  $m_{ini}$ denotes the initial mass, the time-varying mass ratio is 
\begin{equation}
    m_R(t) = \cfrac{m(t)}{m_{ini}},
\end{equation}
\noindent The final mass ratio $m_{R,fin}$ (at $t_{fin}$) is an unknown parameter, as well as the overall durations, $\Delta t_{ini}$ and $\Delta t_{fin}$, of the two thrust arcs. 

The algorithm is initialized by setting $m_{R,fin}$ to the guess value provided by the Tsiolkowsky law,
\begin{equation}\label{MRfinGuess}
    m_{R,fin} = \exp\left[-\frac{\Delta v}{c}\right],
\end{equation} 
where $\Delta v$ is the total velocity change associated with the two-impulse transfer. Moreover, two (initial) boundary times and states are defined, associated with the spacecraft position and velocity at $t_{ini}$ and $t_{fin}$:
\begin{itemize}
    \item $t_{ini}^{(0)}=t_{ini}$, $\boldsymbol{r}_{ini}^{(0)}=\boldsymbol{r}_{ini}$, $\boldsymbol{v}_{ini}^{(0)}=\boldsymbol{v}_{ini}$;
    \item $t_{fin}^{(0)}=t_{fin}$, $\boldsymbol{r}_{fin}^{(0)}=\boldsymbol{r}_{fin}$, $\boldsymbol{v}_{fin}^{(0)}=\boldsymbol{v}_{fin}$;
\end{itemize}
\noindent 

\noindent A sequence of Lambert solutions are sought, starting from the initial boundary times and states (superscript 0). Updated Lambert boundary conditions, associated with increasing indices as superscripts, i.e., $\left\{t_{ini}^{(k)},\boldsymbol{r}_{ini}^{(k)},\boldsymbol{v}_{ini}^{(k)}\right\}$ and $\left\{t_{fin}^{(k)},\boldsymbol{r}_{fin}^{(k)},\boldsymbol{v}_{fin}^{(k)}\right\}$, are obtained through the steps that follow. 
\begin{enumerate}[start=1]
    \item Find the Lambert solution connecting $\boldsymbol{r}_{ini}^{(k)}$ to $\boldsymbol{r}_{fin}^{(k)}$ in $\left(t_{fin}^{(k)}-t_{ini}^{(k)}\right)$, with velocity changes $\Delta \boldsymbol{v}_{ini}^{(k)}$ and $\Delta \boldsymbol{v}_{fin}^{(k)}$, at times $t_{ini}^{(k)}$ and $t_{fin}^{(k)}$, respectively.
    \item Calculate the total velocity change $\Delta v^{(k)}=\Delta v_{ini}^{(k)}+\Delta v_{fin}^{(k)}$, with $\Delta v_{ini/fin}^{(k)}=\left|\Delta \boldsymbol{v}_{ini/fin}^{(k)}\right|$. If $\Delta v^{(k)}<\Delta v_{th}$, then stop the process, otherwise go to Step 3; $\Delta v_{th}$ denotes a modest (threshold) value.
    \item Evaluate the (hypothetical) ignition times for the finite thrust along the initial and final arc, associated with $\Delta \boldsymbol{v}_{ini}^{(k)}$ and $\Delta \boldsymbol{v}_{fin}^{(k)}$, under the approximating assumption of piecewise constant thrust acceleration in each interval, i.e.
    \begin{equation}
        \Delta t_{ini,U}^{(k)} = \frac{ {\Delta v}_{ini}^{(k)}m_{ini}^{(k)}}{T}, \quad\quad\quad \Delta t_{fin,U}^{(k)} = \frac{ {\Delta v}_{fin}^{(k)} m_{fin}^{(k)}}{T}
    \end{equation}
    \noindent where $m_{ini}^{(k)}$ and $m_{fin}^{(k)}$ denote the spacecraft mass at $t_{ini}^{(k)}$ and $t_{fin}^{(k)}$, respectively.
    \item Select the maximum value between $\Delta t_{ini,U}^{(k)}$ and $\Delta t_{fin,U}^{(k)}$, and saturate the corresponding duration to the maximum allowed value $\Delta t_{max}$,
    \begin{equation}
        \Delta t_{j}^{(k)} = \min\left\{\Delta t_{j,U}^{(k)},\Delta t_{max}\right\},
        \quad \textrm{with}\quad j=\arg\max_{l \in C}
        \Delta t_{l,U}^{(k)},\,\,C=\{ini,fin\} 
    \end{equation}
    \item Calculate the remaining interval
    \begin{equation}
        \Delta t_{m}^{(k)} = \Delta t_{m,U}^{(k)}\frac{\Delta t_{j}^{(k)}}{\Delta t_{j,U}^{(k)}}, \quad \textrm{with}\quad m\in C,\,\,m\neq j         
    \end{equation}    
    \item Simultaneously propagate the state (i.e., position and velocity) along the two thrust arcs, using spherical coordinates \cite{Pontani2023}, with time-varying mass in the instantaneous thrust acceleration; more specifically, perform
    \begin{itemize}
        \item  forward propagation in $\left[ t_{ini}^{(k)}, t_{ini}^{(k)} + \Delta t_{ini}^{(k)} \right]$, while assuming thrust aligned with $\Delta \boldsymbol{v}_{ini}^{(k)}$, and
        \item  backward propagation in $\left[ t_{fin}^{(k)}-\Delta t_{fin}^{(k)},  t_{fin}^{(k)} \right]$, while assuming thrust aligned with $\Delta \boldsymbol{v}_{fin}^{(k)}$
    \end{itemize}    
    \item Update the Lambert boundary conditions, to get 
    \begin{itemize}
        \item $\left\{t_{ini}^{(k+1)},\boldsymbol{r}_{ini}^{(k+1)},\boldsymbol{v}_{ini}^{(k+1)}\right\}$, where $t_{ini}^{(k+1)}=t_{ini}^{(k)}+\Delta t_{ini}^{(k)}$, while the updated position and velocity represent the state attained at the end of forward propagation;
        \item $\left\{t_{fin}^{(k+1)},\boldsymbol{r}_{fin}^{(k+1)},\boldsymbol{v}_{fin}^{(k+1)}\right\}$, where $t_{fin}^{(k+1)}=t_{fin}^{(k)}-\Delta t_{fin}^{(k)}$, while the updated position and velocity represent the state attained at the end of backward propagation.
    \end{itemize}
    \item Repeat steps 1 through 7.
\end{enumerate}
The preceding steps lead to finding decreasing values of $\Delta t_{ini}^{(k)}$ and $\Delta t_{fin}^{(k)}$ as $k$ increases. In Step 2, symbol $\Delta v_{th}$ represents a modest threshold value. Under the assumption that the inequality at Step 2 is met when $k=n$, a negligible total velocity change $\Delta v^{(n)}$ would be needed to adjust the trajectory and obtain an elliptic transfer arc connecting $\boldsymbol{r}_{ini}^{(n)}$ to $\boldsymbol{r}_{fin}^{(n)}$. If this occurs, then the process stops, and the interval $\left[t_{ini}^{(n)},t_{fin}^{(n)}\right]$ is recognized as a coast arc, whereas $\left[t_{ini}^{(0)},t_{ini}^{(n)}\right]$ and $\left[t_{fin}^{(n)},t_{fin}^{(0)}\right]$ represent the two (initial and final) thrust arcs. Strictly speaking, the thrust pointing angle is piecewise constant, because the thrust is aligned with the direction of $\Delta \boldsymbol{v}_{ini}^{(k)}$ (or $\Delta \boldsymbol{v}_{fin}^{(k)}$) in each interval $\left[t_{ini}^{(k)},t_{ini}^{(k)}+\Delta t_{ini}^{(k)}\right]$ (or $\left[t_{fin}^{(k)}-\Delta t_{fin}^{(k)},t_{fin}^{(k)}\right]$). Since the latter thrust intervals have maximum duration equal to $\Delta t_{max}$ (cf. Step 4), the choice of a sufficiently small value for $\Delta t_{max}$ has two advantages: (i) the piecewise pointing angle is subject to limited discontinuities and (ii) its overall time history approaches a smooth (continuous) behavior. However, reducing $\Delta t_{max}$ increases the runtime of the algorithm, therefore a judicious choice of $\Delta t_{max}$ must come as a compromise between computational efficiency and continuity of the thrust direction time history.

The previous algorithmic steps need a specific value for the unknown mass ratio $m_{R,fin}$. After completing Steps 1 through 8, the mass ratios $m_{R,ini}^{(n)}$ and $m_{R,fin}^{(n)}$ (at $t_{ini}^{(n)}$ and $t_{fin}^{(n)}$) are obtained, and they must match, because no thrust is used along the (unpowered) coast arc $\left[t_{ini}^{(n)},t_{fin}^{(n)}\right]$. This circumstance implies the need of repeating the preceding steps 1 though 8 several times, while adjusting $m_{R,fin}$, until the following convergence criterion is met:
\begin{equation}
        \left|m_{R,ini}^{(n)}-m_{R,fin}^{(n)}\right|<\epsilon
\end{equation}
where $\epsilon$ represents a prescribed (modest) tolerance on the mass ratio mismatch. 

\begin{figure*}
    \centering
    \includegraphics[width=0.95\textwidth]{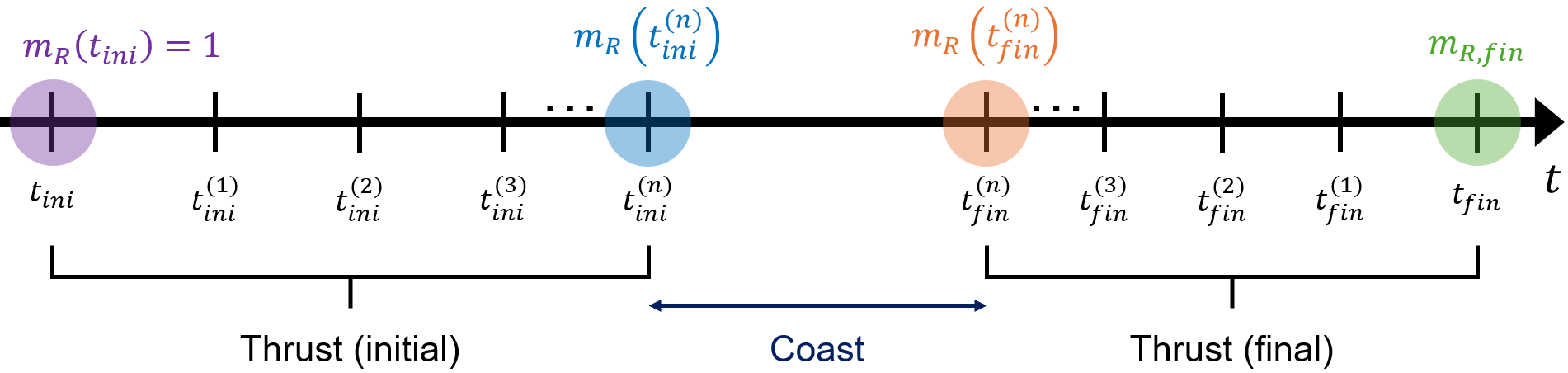}
    \vspace*{-3.5mm}
    \caption{Illustrative sketch of the inner loop (Steps 1 through 8) of the finite-thrust Lambert solver.}
    \label{FIG:LTLS}
\end{figure*}

In summary, the conversion algorithm is composed of an inner loop, i.e. Steps 1 through 8, and an outer loop, which iterates the inner loop several times, with the objective of refining the final mass ratio $m_{R,fin}$, starting from the initial guess provided by Eq. \eqref{MRfinGuess}. An illustrative sketch of the inner loop is provided in Fig. \ref{FIG:LTLS}, which portrays the thrust and coast arcs in relation to the sequences $t_{ini}^{(k)}$ and $t_{fin}^{(k)}$ $(k=1,\ldots,n)$, while highlighting $m_R$ at notable points along the transfer. For the outer loop, a nonlinear programming solver based on the Nelder–Mead method is used in this study.

\section{Results}\label{RD}

\indent This section presents and comments on the numerical results obtained with the proposed trajectory design strategy. First, the mission specifications and requirements are delineated. Then, impulsive and low-thrust round-trip transfers are presented and analyzed. 

\subsection{Mission profile and accessibility analysis}\label{MPAA} 
\indent To illustrate the application of the proposed technique, a representative unmanned mission scenario is investigated, associated with the following constraints:\
\begin{align}
    &t_0^{(\mathrm{max})} - t_0^{(\mathrm{min})} = 5 \; \mathrm{year} \text{,}\\
    &\Delta t_S \in \left[ 100 \; \mathrm{day}, \; 3 \; \mathrm{year} \right] \text{,}\\
    &\Delta t \in \left[ 1, \; 5 \right] \; \mathrm{year} \text{.}
\end{align}
\noindent Moreover,
\begin{equation}
    \delta \varpi_{S/C} = 1 \; \mathrm{degree}, \quad \Delta \varpi_{S/C}^{(\mathrm{max})} = 0.01 \; \mathrm{degree}. 
\end{equation}

\indent This selection is informed by the operational profiles of previous sample-return missions. Hayabusa spent between 2 and 3 months in the vicinity of Itokawa, while Hayabusa2 and OSIRIS-REx operated around Ryugu for $\approx$1.5 years and Bennu for $\approx$2.5 years, respectively. These missions motivate the wide range of stay times examined in this analysis, demonstrating that the S/C may remain in close proximity of an asteroid for both brief and extended periods, with the duration dictated by scientific objectives, available launch and return windows, or unexpected events.

\begin{figure*}
    \centering
    \includegraphics[width=0.85\textwidth]{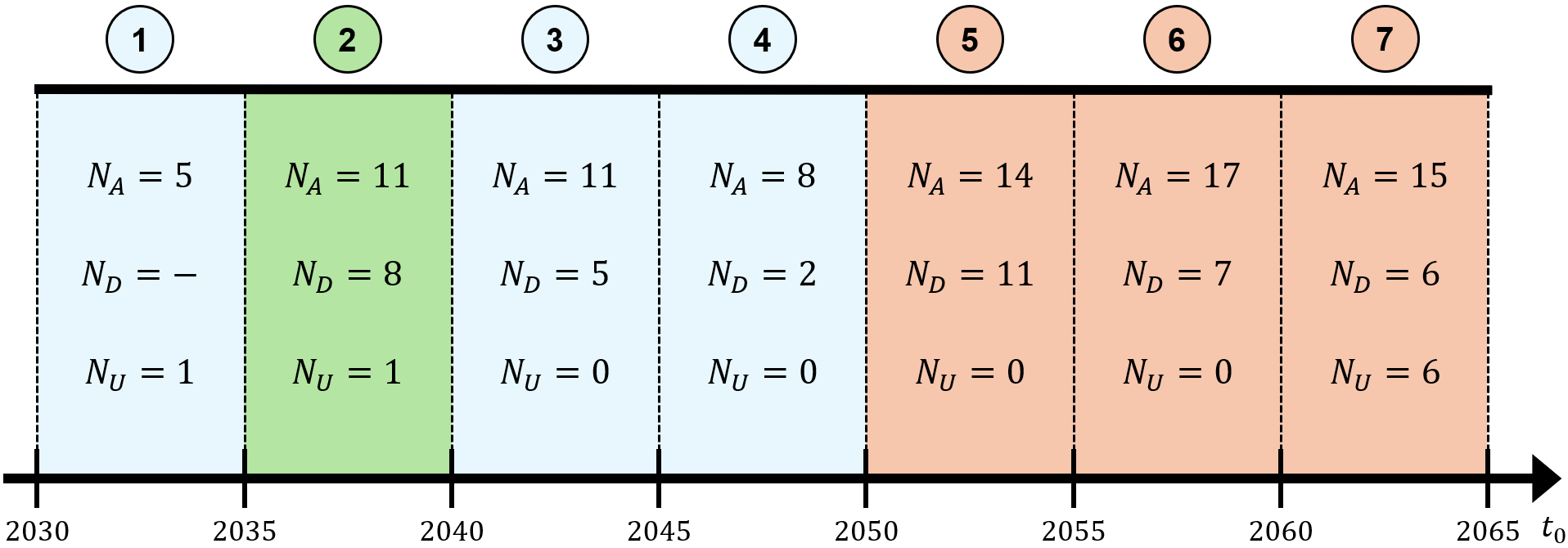}
    \vspace*{-3.5mm}
    \caption{Accessibility analysis: window-specific performance metrics.}
    \label{FIG:ACCAN}
\end{figure*}

\indent Before selecting specific epochs, an accessibility analysis is performed with the objective of identifying a convenient window for the mission. An object is classified as accessible if at least one round-trip solution to that object exists. Multiple adjacent departure windows are evaluated using the following performance metrics:
\begin{itemize}
    \item $N_A :=$ Number of accessed NEOs;
    \item $N_D :=$ Number of accessed NEOs not accessible in the previous window;
    \item $N_U :=$ Number of NEOs accessible only in one window.
\end{itemize}
\noindent The analysis spans the period from 2030-Jan-01 to 2065-Jan-01 and is performed by rigidly shifting a 5-year departure window at 5 year intervals. This preliminary assessment only determines which objects can be accessed in each window, without enforcing the thresholds $\Delta v_2^{(\mathrm{max})}$ and $\Delta v_3^{(\mathrm{max})}$ associated with the one-way transfer procedure presented in Sect. \ref{TDS}. The results of the analysis are summarized in Fig. \ref{FIG:ACCAN}. Because different epochs correspond to different Earth-NEO configurations, shifting the departure window results in different sets of accessible NEOs. Although windows 5, 6, and 7 offer access to a larger pool of objects, windows 2 was preferred due to both its proximity in time and the number of accessed NEOs. Therefore, the selected departure window is $\left[ \text{2035-Jan-1}, \; \text{2040-Jan-1} \right]$, and the corresponding arrival window is $\left[ \text{2036-Jan-1}, \; \text{2045-Jan-1} \right]$. Note that if not specified otherwise, all epochs are 00:00:00 UTC. Moreover, $\Delta v_2^{(\mathrm{max})} = \Delta v_3^{(\mathrm{max})} = 1.5\ \text{km/s}$.\\ 

\subsection{Impulsive round-trip transfers}\label{IRTT}

\begin{table}
\centering
{\renewcommand{\arraystretch}{1.1}
\footnotesize
\caption{Accessible NEO indices and number of one-way transfers classified in terms of gate and transfer type.}
\label{TABOW}
\begin{tabularx}{0.5\linewidth}{|c|X|X|X|X|}
\hline
\makecell{NEO\\ Indices} &
\multicolumn{4}{c|}{
\makecell{
$\left\{ 2,10,12,18,20,23,26,28,29,35,42, \right.$\\
$\left.46,50,55,59,62,64,65,66,70,76,79 \right\}$
}}\\
\hline
 & $L_1$ MT & $L_1$ TO & $L_2$ MT & $L_2$ TO \\
\cline{2-5}
OUTBOUND & 13242 & 5953 & 39397 & 6408 \\
\hline
INBOUND & 22042 & 6264 & 52344 & 11632 \\
\hline
\end{tabularx}
}
\end{table}

\indent First, outbound legs to candidate NEOs are computed, resulting in a total of 65000 trajectories to a subset of 22 out of 80 objects in the target population. Earth return segments exist for all asteroids in the outbound subset, yielding a total of 92282 inbound legs. Each segment can be classified according to the associated gate ($L_1$ or $L_2$) and transfer type (MT or TO). Table \ref{TABOW} provides information on one-way transfers, indicating the indices of accessible objects and the number of solutions classified in terms of gate and transfer type. 

\begin{table}
\centering
{\renewcommand{\arraystretch}{1.1}
\footnotesize
\caption{Number of accessed NEOs per outbound-inbound combination.}
\label{TABNEOIND}
\begin{tabular}{|c
                |c
                |c
                |c
                |c
                |c
                |c|}
\cline{4-7}
\multicolumn{3}{c|}{} & \multicolumn{4}{c|}{INBOUND} \\
\cline{4-7}
\multicolumn{3}{c|}{} & \multicolumn{2}{c|}{$L_1$} & \multicolumn{2}{c|}{$L_2$} \\
\cline{4-7}
\multicolumn{3}{c|}{} & MT & TO & MT & TO \\
\hline
\multirow{4}{*}{\rotatebox[origin=c]{90}{OUTBOUND}} 
& \multirow{2}{*}{$L_1$} & MT & 4 & 4 & 6 & 5 \\
\cline{3-7}
&                        & TO & 3 & 3 & 4 & 4 \\
\cline{2-7}
& \multirow{2}{*}{$L_2$} & MT & 6 & 6 & 4 & 3 \\
\cline{3-7}
&                        & TO & 6 & 6 & 2 & 2 \\
\hline
\end{tabular}
}
\end{table}

\begin{table}
\centering
{\renewcommand{\arraystretch}{1.1}
\footnotesize
\caption{Number of round-trip solutions per outbound-inbound combination.}
\label{TABTOTRTSOL}
\begin{tabular}{|c
                |c
                |c
                |c
                |c
                |c
                |c|}
\cline{4-7}
\multicolumn{3}{c|}{} & \multicolumn{4}{c|}{INBOUND} \\
\cline{4-7}
\multicolumn{3}{c|}{} & \multicolumn{2}{c|}{$L_1$} & \multicolumn{2}{c|}{$L_2$} \\
\cline{4-7}
\multicolumn{3}{c|}{} & MT & TO & MT & TO \\
\hline
\multirow{4}{*}{\rotatebox[origin=c]{90}{OUTBOUND}} 
& \multirow{2}{*}{$L_1$} & MT & 114419 & 26786 & 743246 & 18055 \\
\cline{3-7}
&                        & TO & 66139 & 11192 & 25659 & 626 \\
\cline{2-7}
& \multirow{2}{*}{$L_2$} & MT & 603880 & 19752 & 139656 & 104293 \\
\cline{3-7}
&                        & TO & 41679 & 17854 & 83090 & 53078 \\
\hline
\end{tabular}
}
\end{table}

\indent Combining outbound and inbound segments yields a total of about 2 million round-trip transfers to 11 ($\left\{ 2,20,26,28,29,35,42,55,59,65,66 \right\}$) of the 22 objects listed in Table \ref{TABOW}. The extensive solution space can be approached by gathering insights on the distribution of round-trip trajectories across the 16 types of outbound-inbound combinations. Tables \ref{TABNEOIND} and \ref{TABTOTRTSOL} present the number of accessed objects and round-trip transfers per outbound-inbound combination, respectively. Table \ref{TABNEOIND} shows that, for the subset of targets considered in this study, alternate-gate (AG) connections (i.e., Earth is left and accessed through different gates) enable access to a larger number of objects with respect to same-gate (SG) connections. The predominance of MT-MT combinations observable in Table \ref{TABTOTRTSOL} arises from the larger number of one-way MT trajectories in the S/C orbit database (see Table \ref{TABOW}) exacerbated by combinatorial explosion. Unlike classical optimization-driven searches, the modular construction of the trajectory database enables a direct link between the underlying dynamical structures and the resulting mission opportunities. Instead of solving an inverse problem to determine suitable launch and return windows, the method simply selects candidate epochs from specified windows, propagates the associated trajectories, and verifies the satisfaction of a limited set of constraints with clear physical meaning, streamlining the design process and improving interpretability. 

\begin{figure*}
    \centering
    \includegraphics[width=0.95\textwidth]{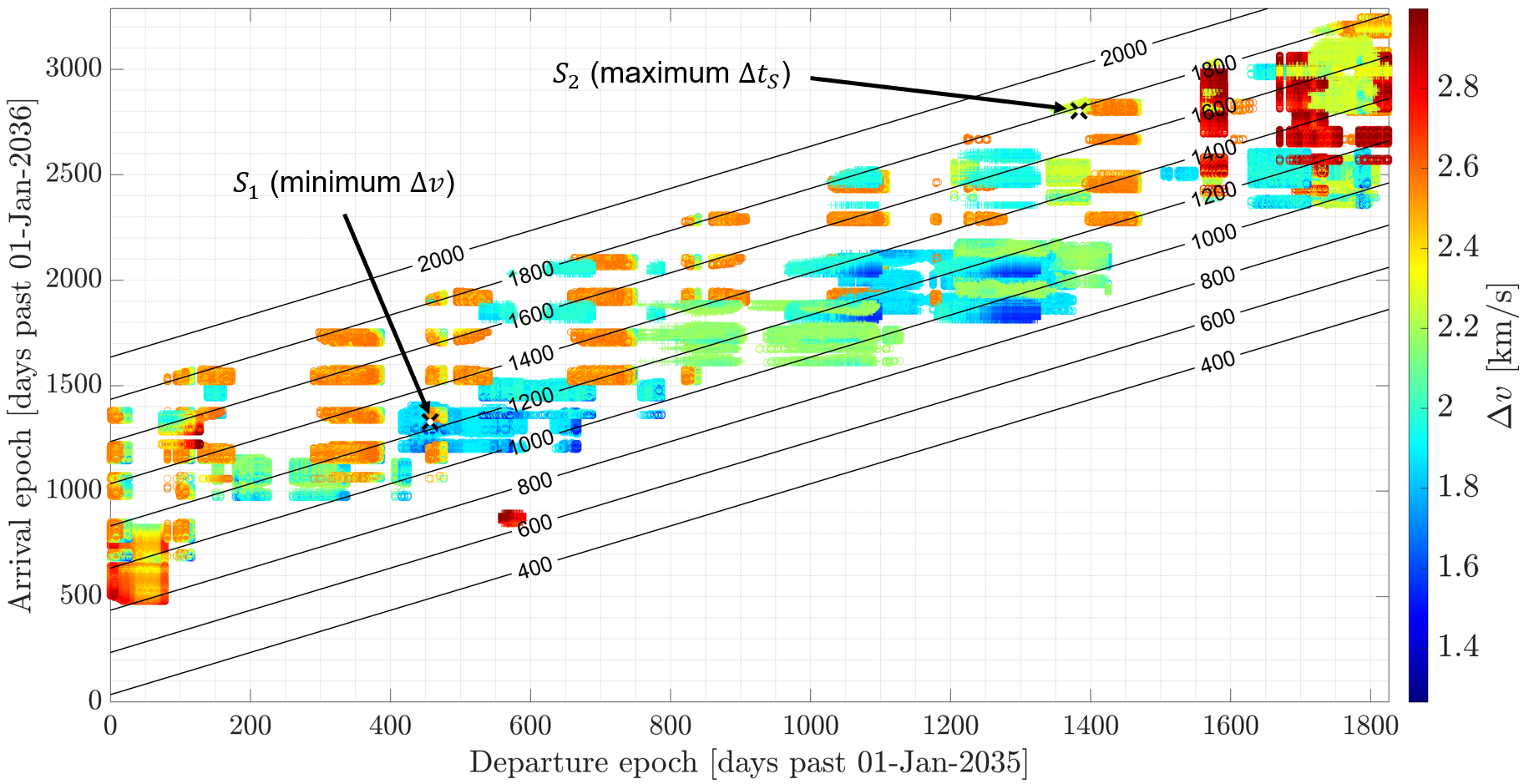}
    \vspace*{-3.5mm}
    \caption{Total impulse against
    departure and arrival dates for impulsive solutions with lines of constant total mission duration in black.}
    \label{FIG:SOLDV}
\end{figure*}

\begin{figure*}
    \centering
    \includegraphics[width=0.95\textwidth]{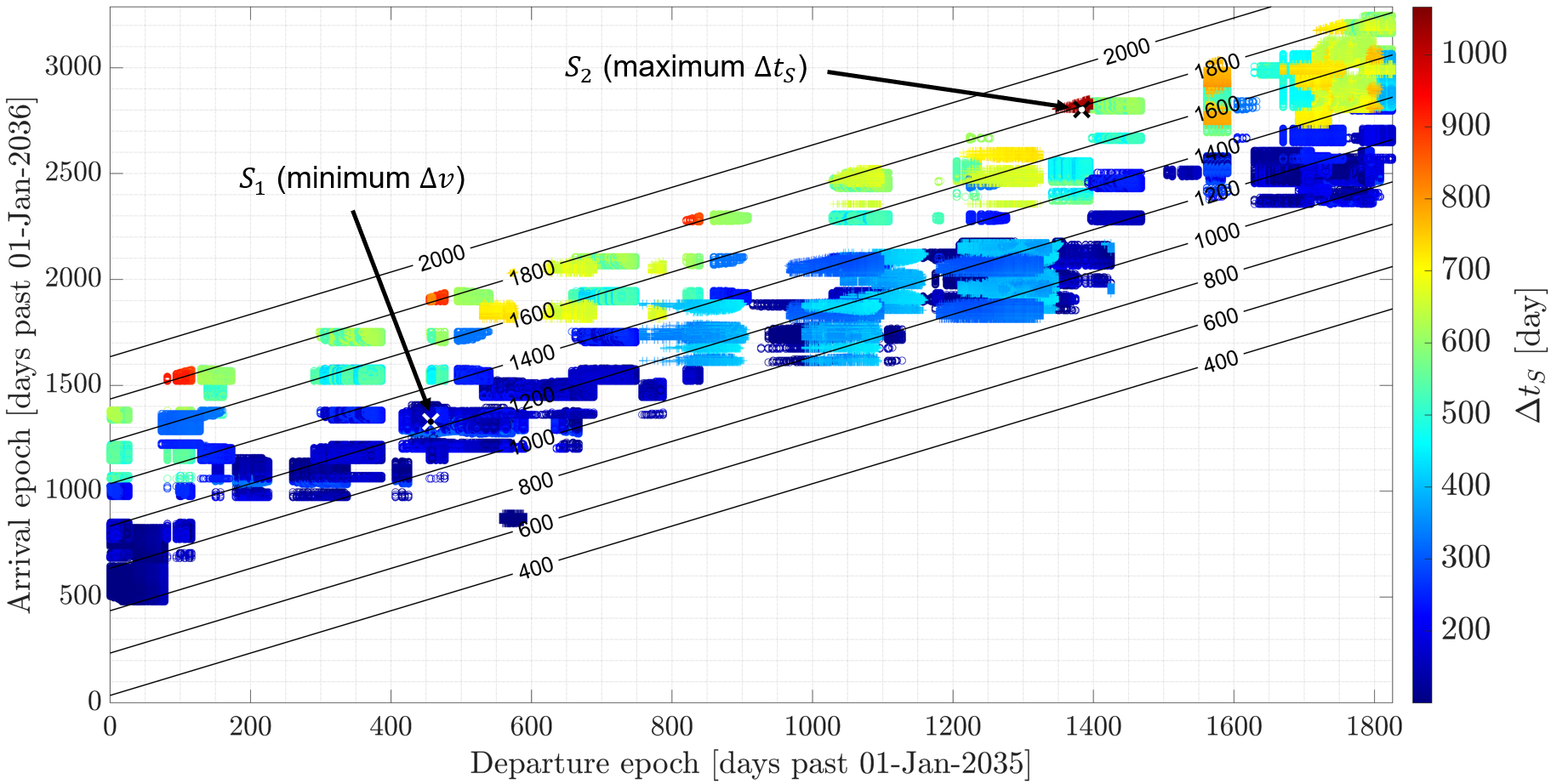}
    \vspace*{-3.5mm}
    \caption{Stay time against
    departure and arrival dates for impulsive solutions with lines of constant total mission duration in black.}
    \label{FIG:SOLDTS}
\end{figure*}

\begin{figure*}
    \centering
    \includegraphics[width=0.95\textwidth]{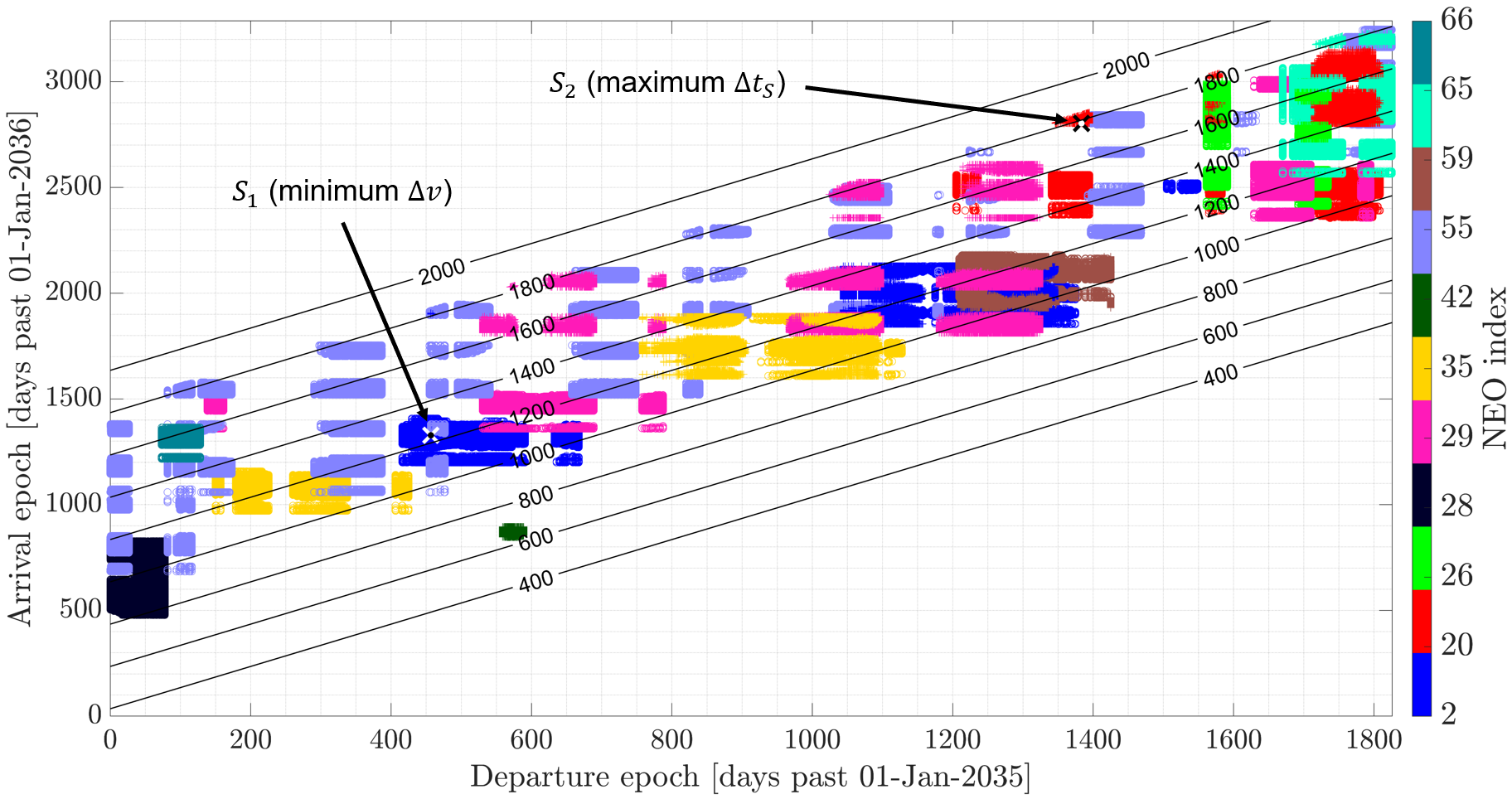}
    \vspace*{-3.5mm}
    \caption{NEO index against
    departure and arrival dates for impulsive solutions with lines of constant total mission duration in black.} 
    \label{FIG:SOLNEOIND}
\end{figure*}

\begin{table*}
\centering
\caption{Characteristics of the best round-trip impulsive transfers under gate-based classification. For each object, the minimum $\Delta v$ and maximum $\Delta t_S$ trajectories are presented in gray and white rows, respectively. The overall minimum $\Delta v$ ($S_1$) and maximum $\Delta t_S$ ($S_2$) solutions are highlighted in light blue and red, respectively.}
\begin{adjustbox}{max width=\textwidth}
{\renewcommand{\arraystretch}{1.0}
\renewcommand{\tabcolsep}{4.00pt} 
\label{BIGTABGATE}
\begin{tabular}{ll|llll|llll|llll|llll}
\multirow{3}{*}{\makecell[l]{\textbf{NEO}\\ \textbf{Ind.}}} & \multirow{3}{*}{\makecell[l]{\textbf{No. Sol.}\\ a/b/c/d}} &
\multicolumn{4}{c|}{$L_1$–$L_1$ (a)} &
\multicolumn{4}{c|}{$L_1$–$L_2$ (b)} &
\multicolumn{4}{c|}{$L_2$–$L_1$ (c)} &
\multicolumn{4}{c}{$L_2$–$L_2$ (d)} \\
\cline{3-18} 
& & \multirow{2}{*}{\makecell{$\Delta v$\\ $[\nicefrac{\text{km}}{\text{s}}]$}} & \multirow{2}{*}{\makecell{$\Delta t_{TC}$\\ $[\text{day}]$}} & \multirow{2}{*}{\makecell{$\Delta t_{H}$\\ $[\text{day}]$}} & \multirow{2}{*}{\makecell{$\Delta t_{S}$\\ $[\text{day}]$}} &
\multirow{2}{*}{\makecell{$\Delta v$\\ $[\nicefrac{\text{km}}{\text{s}}]$}} & \multirow{2}{*}{\makecell{$\Delta t_{TC}$\\ $[\text{day}]$}} & \multirow{2}{*}{\makecell{$\Delta t_{H}$\\ $[\text{day}]$}} & \multirow{2}{*}{\makecell{$\Delta t_{S}$\\ $[\text{day}]$}} &
\multirow{2}{*}{\makecell{$\Delta v$\\ $[\nicefrac{\text{km}}{\text{s}}]$}} & \multirow{2}{*}{\makecell{$\Delta t_{TC}$\\ $[\text{day}]$}} & \multirow{2}{*}{\makecell{$\Delta t_{H}$\\ $[\text{day}]$}} & \multirow{2}{*}{\makecell{$\Delta t_{S}$\\ $[\text{day}]$}} &
\multirow{2}{*}{\makecell{$\Delta v$\\ $[\nicefrac{\text{km}}{\text{s}}]$}} & \multirow{2}{*}{\makecell{$\Delta t_{TC}$\\ $[\text{day}]$}} & \multirow{2}{*}{\makecell{$\Delta t_{H}$\\ $[\text{day}]$}} & \multirow{2}{*}{\makecell{$\Delta t_{S}$\\ $[\text{day}]$}} \\
 &  &  &  &  &  &  &  &  &  &  &  &  &  &  &  &  & \\
\hline

\rowcolor[gray]{0.9}
\multirow{2}{*}{\textbf{2}}  & \multirow{2}{*}{\makecell[l]{66948/284193/\\34532/0}}&1.616&565.2&413.5&406.2&\cellcolor{LightBlue1}\textbf{1.265}&\cellcolor{LightBlue1}535.7&\cellcolor{LightBlue1}449.9&\cellcolor{LightBlue1}250.6&1.627&569.7&550.9&141.5&-&-&-&-\\
& &1.893&300.6&695.0&813.3&1.890&588.8&332.8&336.2&1.895&342.2&465.8&207.2&-&-&-&-\\

\hline

\rowcolor[gray]{0.9}
\multirow{2}{*}{\textbf{20}} & \multirow{2}{*}{\makecell[l]{0/0/\\155694/116431}}&-&-&-&-&-&-&-&-&1.349&564.7&391.1&166.1&1.699&347.3&411.5&678.1\\
& &-&-&-&-&-&-&-&-&2.294&586.2&437.2&653.4&\cellcolor{LightRed1}2.298&\cellcolor{LightRed1}319.9&\cellcolor{LightRed1}398.2&\cellcolor{LightRed1}\textbf{1066.4}\\

\hline

\rowcolor[gray]{0.9}
\multirow{2}{*}{26} & \multirow{2}{*}{\makecell[l]{21631/47453/\\0/0}}&2.509&595.6&326.0&708.1&1.737&564.5&484.8&118.3&-&-&-&-&-&-&-&-\\
& &2.958&540.9&378.5&817.6&2.978&554.0&445.0&608.2&-&-&-&-&-&-&-&-\\

\hline

\rowcolor[gray]{0.9}
\multirow{2}{*}{28} & \multirow{2}{*}{\makecell[l]{0/0/\\37314/0}}&-&-&-&-&-&-&-&-&2.056&416.4&625.7&111.7&-&-&-&-\\
& &-&-&-&-&-&-&-&-&2.917&452.3&429.3&182.0&-&-&-&-\\

\hline

\rowcolor[gray]{0.9}
\multirow{2}{*}{29} & \multirow{2}{*}{\makecell[l]{0/84799/\\48835/249625}}&-&-&-&-&1.348&589.3&494.4&129.6&1.380&569.0&576.4&170.6&1.501&550.1&479.1&347.8\\
& &-&-&-&-&1.998&461.9&526.1&587.4&1.999&619.3&546.5&569.2&1.986&373.1&567.4&751.9\\

\hline

\rowcolor[gray]{0.9}
\multirow{2}{*}{35} & \multirow{2}{*}{\makecell[l]{103080/123123/\\14211/0}}&1.816&568.8&345.4&369.8&1.480&554.8&454.1&260.3&1.922&564.1&474.3&102.2&-&-&-&-\\
& &2.191&342.9&367.7&448.9&2.194&590.4&343.1&342.8&2.193&567.4&489.5&193.3&-&-&-&-\\

\hline

\rowcolor[gray]{0.9}
\multirow{2}{*}{42} & \multirow{2}{*}{\makecell[l]{0/0/\\0/5283}}&-&-&-&-&-&-&-&-&-&-&-&-&2.443&579.9&3.3&100.1\\
& &-&-&-&-&-&-&-&-&-&-&-&-&2.998&573.9&17.4&109.8\\

\hline

\rowcolor[gray]{0.9}
\multirow{2}{*}{55} & \multirow{2}{*}{\makecell[l]{0/0/\\338169/0}}&-&-&-&-&-&-&-&-&1.500&517.0&521.1&100.4&-&-&-&-\\
& &-&-&-&-&-&-&-&-&2.597&399.6&470.2&922.0&-&-&-&-\\

\hline

\rowcolor[gray]{0.9}
\multirow{2}{*}{59} & \multirow{2}{*}{\makecell[l]{26877/240426/\\0/0}}&1.967&535.2&342.0&401.3&1.661&526.1&540.4&149.2&-&-&-&-&-&-&-&-\\
& &2.189&410.2&340.0&422.0&2.195&441.8&426.2&253.8&-&-&-&-&-&-&-&-\\

\hline

\rowcolor[gray]{0.9}
\multirow{2}{*}{65} & \multirow{2}{*}{\makecell[l]{0/0/\\54410/8778}}&-&-&-&-&-&-&-&-&1.689&555.6&606.4&148.0&2.080&342.9&560.4&726.8\\
& &-&-&-&-&-&-&-&-&2.967&589.3&502.6&525.3&2.463&548.0&513.6&755.9\\

\hline

\rowcolor[gray]{0.9}
\multirow{2}{*}{66} & \multirow{2}{*}{\makecell[l]{0/7592/\\0/0}}&-&-&-&-&1.862&590.2&730.5&300.7&-&-&-&-&-&-&-&-\\
& &-&-&-&-&2.607&612.5&672.9&325.6&-&-&-&-&-&-&-&-\\

\end{tabular}
}
\end{adjustbox}
\end{table*}

\begin{table*}
\centering
\caption{Characteristics of the best round-trip impulsive transfers under transfer type-based classification. For each object, the minimum $\Delta v$ and maximum $\Delta t_S$ trajectories are presented in gray and white rows, respectively. The overall minimum $\Delta v$ ($S_1$) and maximum $\Delta t_S$ ($S_2$) solutions are highlighted in light blue and red, respectively.}
\begin{adjustbox}{max width=\textwidth}
{\renewcommand{\arraystretch}{1.0}
\renewcommand{\tabcolsep}{4.00pt} 
\label{BIGTABTTYPE}
\begin{tabular}{ll|llll|llll|llll|llll}
\multirow{3}{*}{\makecell[l]{\textbf{NEO}\\ \textbf{Ind.}}} & \multirow{3}{*}{\makecell[l]{\textbf{No. Sol.}\\ a/b/c/d}} &
\multicolumn{4}{c|}{MT-MT (a)} &
\multicolumn{4}{c|}{MT-TO (b)} &
\multicolumn{4}{c|}{TO-MT (c)} &
\multicolumn{4}{c}{TO-TO (d)} \\
\cline{3-18} 
& & \multirow{2}{*}{\makecell{$\Delta v$\\ $[\nicefrac{\text{km}}{\text{s}}]$}} & \multirow{2}{*}{\makecell{$\Delta t_{TC}$\\ $[\text{day}]$}} & \multirow{2}{*}{\makecell{$\Delta t_{H}$\\ $[\text{day}]$}} & \multirow{2}{*}{\makecell{$\Delta t_{S}$\\ $[\text{day}]$}} &
\multirow{2}{*}{\makecell{$\Delta v$\\ $[\nicefrac{\text{km}}{\text{s}}]$}} & \multirow{2}{*}{\makecell{$\Delta t_{TC}$\\ $[\text{day}]$}} & \multirow{2}{*}{\makecell{$\Delta t_{H}$\\ $[\text{day}]$}} & \multirow{2}{*}{\makecell{$\Delta t_{S}$\\ $[\text{day}]$}} &
\multirow{2}{*}{\makecell{$\Delta v$\\ $[\nicefrac{\text{km}}{\text{s}}]$}} & \multirow{2}{*}{\makecell{$\Delta t_{TC}$\\ $[\text{day}]$}} & \multirow{2}{*}{\makecell{$\Delta t_{H}$\\ $[\text{day}]$}} & \multirow{2}{*}{\makecell{$\Delta t_{S}$\\ $[\text{day}]$}} &
\multirow{2}{*}{\makecell{$\Delta v$\\ $[\nicefrac{\text{km}}{\text{s}}]$}} & \multirow{2}{*}{\makecell{$\Delta t_{TC}$\\ $[\text{day}]$}} & \multirow{2}{*}{\makecell{$\Delta t_{H}$\\ $[\text{day}]$}} & \multirow{2}{*}{\makecell{$\Delta t_{S}$\\ $[\text{day}]$}} \\
 &  &  &  &  &  &  &  &  &  &  &  &  &  &  &  &  & \\
\hline

\rowcolor[gray]{0.9}
\multirow{2}{*}{\textbf{2}}  & \multirow{2}{*}{\makecell[l]{317235/24025/\\35674/8739}}&\cellcolor{LightBlue1} \textbf{1.265}&\cellcolor{LightBlue1} 535.7&\cellcolor{LightBlue1} 449.9&\cellcolor{LightBlue1} 250.6&1.432&410.8&477.6&224.0&1.333&432.8&390.7&202.2&1.500&307.9&418.3&175.6\\
& &1.899&605.6&354.8&437.9&1.893&300.6&695.0&813.3&1.888&358.2&350.6&438.9&1.898&123.3&352.0&440.2\\

\hline

\rowcolor[gray]{0.9}
\multirow{2}{*}{\textbf{20}} & \multirow{2}{*}{\makecell[l]{211626/54422/\\5690/387}}&1.349&564.7&391.1&166.1&1.502&456.8&420.5&141.1&1.798&328.8&489.6&245.1&1.951&220.9&518.9&220.2\\
& &2.293&546.6&491.1&782.7&\cellcolor{LightRed1} 2.298&\cellcolor{LightRed1} 319.9&\cellcolor{LightRed1} 398.2&\cellcolor{LightRed1} \textbf{1066.4}&2.295&337.0&510.5&788.1&2.300&173.8&507.3&788.1\\

\hline

\rowcolor[gray]{0.9}
\multirow{2}{*}{26} & \multirow{2}{*}{\makecell[l]{61596/7488/\\0/0}}&1.737&564.5&484.8&118.3&2.195&465.4&435.2&176.4&-&-&-&-&-&-&-&-\\
& &2.958&540.9&378.5&817.6&2.818&318.1&418.3&787.8&-&-&-&-&-&-&-&-\\

\hline

\rowcolor[gray]{0.9}
\multirow{2}{*}{28} & \multirow{2}{*}{\makecell[l]{0/0/\\20370/16944}}&-&-&-&-&-&-&-&-&2.056&416.4&625.7&111.7&2.095&187.1&625.6&114.7\\
& &-&-&-&-&-&-&-&-&2.917&452.3&429.3&182.0&2.921&221.6&429.7&181.6\\

\hline

\rowcolor[gray]{0.9}
\multirow{2}{*}{29} & \multirow{2}{*}{\makecell[l]{191618/53337/\\85478/52826}}&1.348&589.3&494.4&129.6&1.384&454.9&482.6&137.9&1.392&441.0&477.6&146.8&1.429&306.7&465.8&155.1\\
& &1.996&522.7&560.2&741.9&1.986&373.1&567.4&751.9&1.997&364.2&597.2&750.0&1.996&227.2&499.3&699.8\\

\hline

\rowcolor[gray]{0.9}
\multirow{2}{*}{35} & \multirow{2}{*}{\makecell[l]{185457/6263/\\45292/3402}}&1.480&554.8&454.1&260.3&1.667&427.9&521.8&201.2&1.571&443.1&371.1&215.2&1.758&316.3&438.7&156.0\\
& &2.199&574.6&369.5&447.4&2.198&340.5&383.8&446.4&2.191&342.9&367.7&448.9&2.198&118.1&375.9&448.7\\

\hline

\rowcolor[gray]{0.9}
\multirow{2}{*}{42} & \multirow{2}{*}{\makecell[l]{5283/0/\\0/0}}&2.443&579.9&3.3&100.1&-&-&-&-&-&-&-&-&-&-&-&-\\
& &2.998&573.9&17.4&109.8&-&-&-&-&-&-&-&-&-&-&-&-\\

\hline

\rowcolor[gray]{0.9}
\multirow{2}{*}{55} & \multirow{2}{*}{\makecell[l]{323702/1220/\\13174/73}}&1.500&517.0&521.1&100.4&1.504&411.5&511.7&104.9&1.515&461.8&506.2&107.5&1.516&320.3&506.0&107.6\\
& &2.593&525.7&578.9&639.6&2.586&417.5&613.4&610.6&2.597&399.6&470.2&922.0&2.593&313.4&462.0&556.6\\

\hline

\rowcolor[gray]{0.9}
\multirow{2}{*}{59} & \multirow{2}{*}{\makecell[l]{241748/14287/\\10889/379}}&1.661&526.1&540.4&149.2&1.752&401.4&571.4&132.9&1.727&442.8&494.1&113.4&1.844&305.7&502.7&113.5\\
& &2.174&548.8&315.3&420.5&2.185&342.8&318.7&417.6&2.189&410.2&340.0&422.0&2.194&198.1&346.2&418.3\\

\hline

\rowcolor[gray]{0.9}
\multirow{2}{*}{65} & \multirow{2}{*}{\makecell[l]{55344/7844/\\0/0}}&1.689&555.6&606.4&148.0&2.080&342.9&560.4&726.8&-&-&-&-&-&-&-&-\\
& &2.463&548.0&513.6&755.9&2.386&401.9&525.4&752.4&-&-&-&-&-&-&-&-\\

\hline

\rowcolor[gray]{0.9}
\multirow{2}{*}{66} & \multirow{2}{*}{\makecell[l]{7592/0/\\0/0}}&1.862&590.2&730.5&300.7&-&-&-&-&-&-&-&-&-&-&-&-\\
& &2.607&612.5&672.9&325.6&-&-&-&-&-&-&-&-&-&-&-&-\\

\end{tabular}
}
\end{adjustbox}
\end{table*}

\indent The solutions are visualized through charts inspired by pork chop plots. Figures \ref{FIG:SOLDV}-\ref{FIG:SOLNEOIND} show $\Delta v$, $\Delta t_S$, and NEO index against departure and arrival dates. AG trajectories are denoted by a circle ($\circ$), while a plus symbol (+) indicates SG transfers. Because data points are tightly clustered, readers are referred to Tables~\ref{TABNEOIND} and~\ref{TABTOTRTSOL} for complete gate-related information. The solution space is densely populated by trajectories that cover a wide spectrum of performance levels. Figure \ref{FIG:SOLNEOIND} shows that several objects can be accessed multiple times within the same window, facilitating mission planning and improving flexibility. The identification of multiple viable round-trip transfers to each NEO across a wide time frame, corresponding to different launch and return dates, enables the staggered deployment of multiple S/C to the same asteroid. Alternatively, a single S/C could visit the same NEO or a sequence of different targets across multiple exploration campaigns.\\ 
\indent Although transfers to most asteroids can be found lumped in multiple large compact blocks scattered across the solution space, round-trip opportunities to objects 2017 HU49 (index 29) and 2022 OB5 (index 55) are distributed somewhat regularly across departure and arrival epochs. In contrast, NEOs 2017 FT102 (index 28), 2020 RB4 (index 42), and 2023 RX1 (index 66) offer a limited number of round-trip trajectories enclosed within narrow time frames. The most accessible object (index 55) is also the one that exhibits the largest synodic period with Earth (276.3 years). Tables \ref{BIGTABGATE} and \ref{BIGTABTTYPE} report the minimum $\Delta v$ and maximum $\Delta t_S$ solutions for each object. For some NEOs, certain gate or transfer type combinations do not admit round-trip transfers (dashes in Tables \ref{BIGTABGATE} and \ref{BIGTABTTYPE}).\\  
\indent The best transfers according to the aforementioned criteria are designated with $S_1$ and $S_2$ and highlighted in Figs. \ref{FIG:SOLNEOIND}-\ref{FIG:2SOLLT} and Tables \ref{BIGTABGATE}, \ref{BIGTABTTYPE}, \ref{TABPERFIab}, and \ref{TABPERFLTab}. $S_1$ is a $L_1$ MT–$L_2$ MT transfer to 1991 VG (index 2), a very small NEO of the Apollo group that attracted interest in the past due to its Earth-like orbit and brief capture episodes as a temporary satellite. Because of its accessibility, unusual dynamical behavior, and debated artificial origin, 1991 VG was proposed as a target for NASA's NEA Scout mission. $S_2$ is of the $L_2$ MT–$L_2$ TO class and targets 2014 DJ80 (index 20), the largest object among the 11 accessible NEOs. Note that $S_1$ and $S_2$ represent fundamentally different missions: $S_1$ is an AG trajectory based on MTs, while $S_2$ is an SG transfer with an outbound MT and an inbound TO. This exemplifies the ability of the methodology to identify solutions corresponding to different mission architectures, whether relying on trajectories around PLOs (MTs) or direct paths from/to parking orbits or the Earth's surface (TOs).

\subsection{Low-thrust round-trip transfers}

\begin{table}
\centering
\caption{S/C and engine data used in this work.}
\begin{tblr}{Q[c,m]|Q[c,m]|Q[c,m]|Q[c,m]}
$m_{dry}$ & $T$ & $I_{sp}$ & $P_{in}$ \\
\hline 
600 kg & 240 mN & 1700 s & 4 kW \\
\end{tblr}
\label{TABENG}
\end{table}

\begin{figure*}
    \centering
    \includegraphics[width=0.95\textwidth]{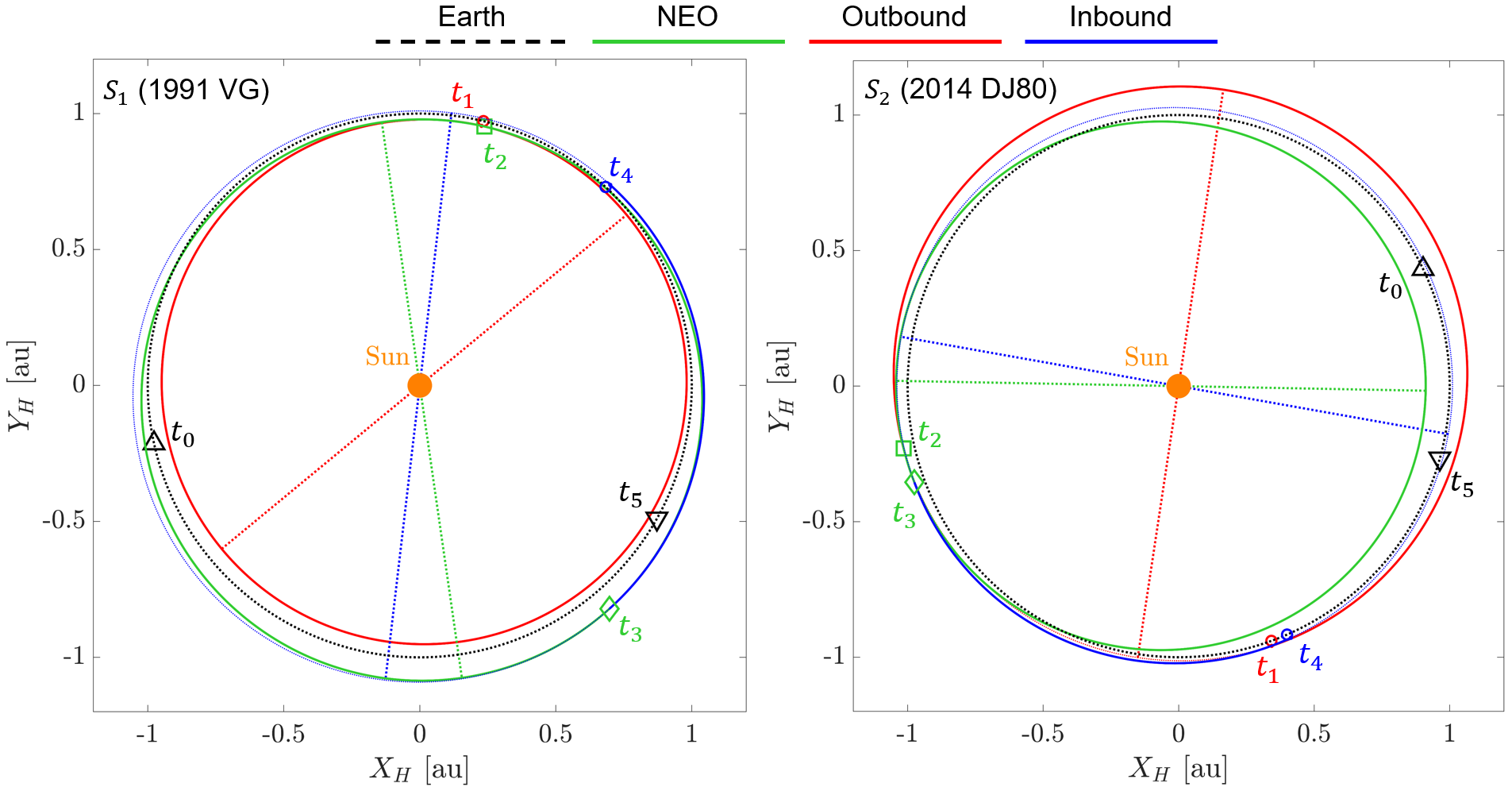}
    \vspace*{-3.5mm}
    \caption{Impulsive round-trip transfers, $S_1$ (left) and $S_2$ (right).} 
    \label{FIG:2SOLI}
\end{figure*}

\begin{figure*}
    \centering
    \includegraphics[width=0.95\textwidth]{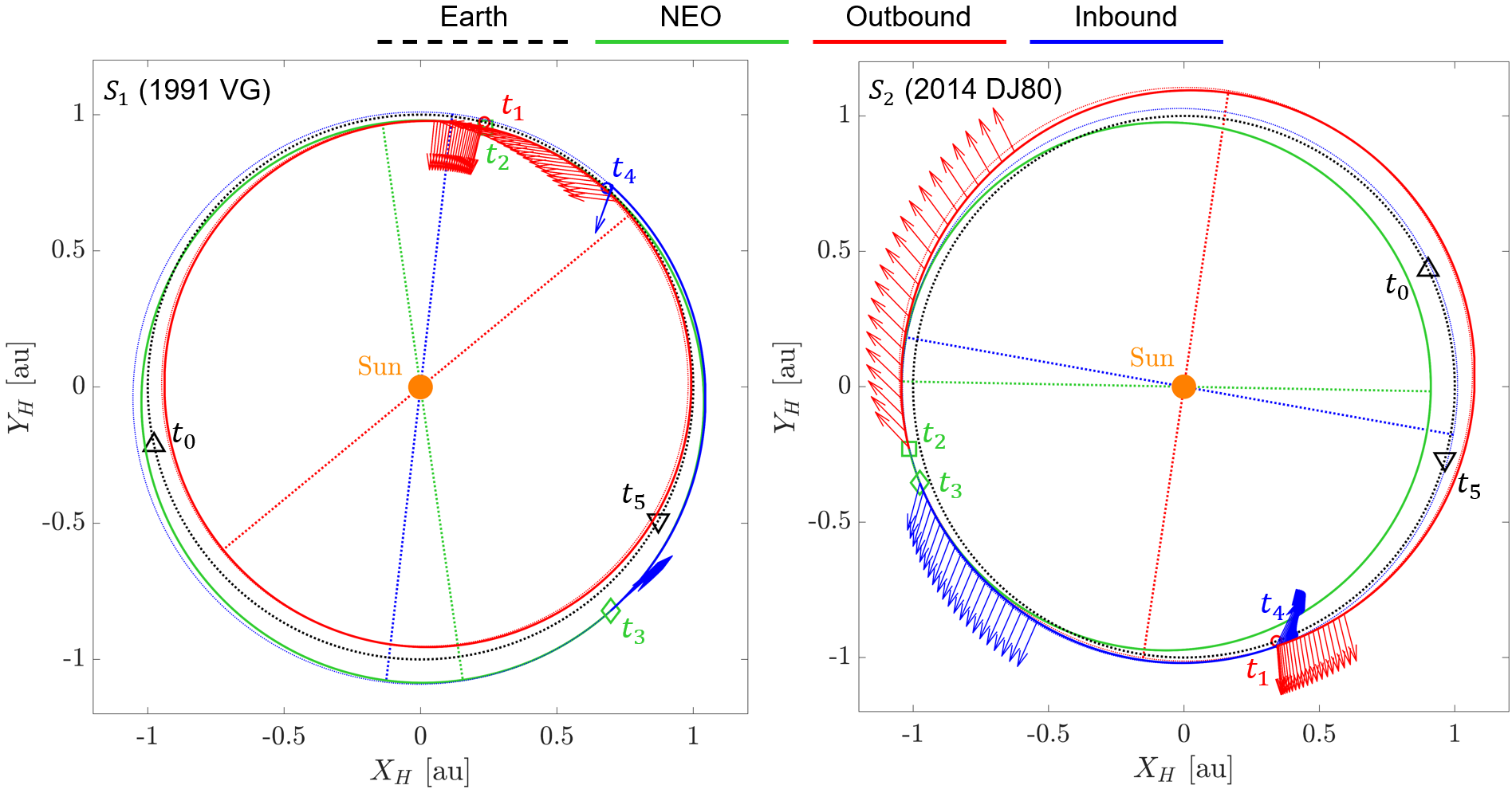}
    \vspace*{-3.5mm}
    \caption{Low-thrust round-trip transfers with arrows showing thrust direction, $S_1$ (left) and $S_2$ (right).} 
    \label{FIG:2SOLLT}
\end{figure*}

\begin{table*}
\centering
\caption{Performance features of impulsive solutions $S_1$ and $S_2$.}
\label{TABPERFIab}
\begin{adjustbox}{max width=\textwidth}
{\renewcommand{\arraystretch}{2.0}
\renewcommand{\tabcolsep}{4.00pt}
\begin{tabular}{c|c|c|c|c|c|c|c|c|c|c|c|c|c}

& $t_0$ & $t_1$ & $t_2$ & $t_3$ & $t_4$ & $t_5$ &
\makecell[c]{$\Delta t_S$\\ $\left[ \mathrm{day} \right]$} &
\makecell[c]{$\Delta t$\\ $\left[ \mathrm{day} \right]$} &
\makecell[c]{$\Delta v_2$\\ $\left[ \nicefrac{\mathrm{m}}{\mathrm{s}} \right]$} &
\makecell[c]{$\Delta v_3$\\ $\left[ \nicefrac{\mathrm{m}}{\mathrm{s}} \right]$} &
\makecell[c]{$C_{3}$\\ $\left[ \nicefrac{\mathrm{km^2}}{\mathrm{s^2}} \right]$} &
\makecell[c]{$v_{\infty}$\\ $\left[ \nicefrac{\mathrm{m}}{\mathrm{s}} \right]$} \\

\hline

\makecell[c]{1991 VG (\textbf{$S_1$})\\$L_1$ MT-$L_2$ MT} &
\makecell[l]{2036-\\Apr-01} & \makecell[l]{2036-\\Dec-05} &
\makecell[l]{2037-\\Nov-16} & \makecell[l]{2038-\\Jul-24} &
\makecell[l]{2038-\\Nov-05} & \makecell[l]{2039-\\Aug-20} &
250.6 & 1236.2 & 1010.3 & 262.4 & 0.239 & 390 \\

\hline

\makecell[c]{2014 DJ80 (\textbf{$S_2$})\\$L_2$ MT-$L_2$ TO} &
\makecell[l]{2038-\\Oct-15} & \makecell[l]{2039-\\Jul-10} &
\makecell[l]{2040-\\May-04} & \makecell[l]{2043-\\Apr-05} &
\makecell[l]{2043-\\Jul-14} & \makecell[l]{2043-\\Sep-03} &
1066.4 & 1784.6 & 1390.5 & 909.3 & 0.014 & 704 \\

\end{tabular}
}
\end{adjustbox}
\end{table*}
\begin{table}
\centering
{\renewcommand{\arraystretch}{1.4}
\renewcommand{\tabcolsep}{4.00pt}
\caption{S/C mass for low-thrust solutions $S_1$ and $S_2$.}
\label{TABPERFLTab}
\begin{tabular}{c|c|c}

& \makecell[c]{$m_0$\\ $\left[ \mathrm{kg} \right]$} & \makecell[c]{$m_2 = m_3$\\ $\left[ \mathrm{kg} \right]$} \\

\hline

\makecell[c]{1991 VG (\textbf{$S_1$})\\$L_1$ MT-$L_2$ MT} & 663.3 & 610.1 \\

\hline

\makecell[c]{2014 DJ80 (\textbf{$S_2$})\\$L_2$ MT-$L_2$ TO} & 765.0 & 656.9 \\

\end{tabular}
}
\end{table}

\indent Because the S/C operates at approximately 1~au from the Sun throughout the mission, solar arrays can provide the power required to operate the low-thrust propulsion system. As an illustrative example, inspired by missions such as Dawn (747-kg dry mass), Hayabusa (490 kg), and DART (610 kg), the S/C is assumed to be equipped with a propulsion system capable of delivering a constant 240 mN thrust at 1700 s specific impulse, operating on 4 kW of input power. These parameters could be nimbly achieved, for example, by clustering two Busek BHT-1500 Hall effect thrusters \cite{BusekBHT1500}. Because the selected targets and low-energy trajectories remain in the vicinity of 1~au, variations on the solar distance and on availability of electric power are limited. Therefore, the assumption of constant propulsion parameters is adequate for preliminary analysis, although more refined SEP models \cite{topputo2021envelop} could be adopted in subsequent development phases of the mission. The proposed S/C-engine configuration is suitable for missions to the asteroids in the target set, typically characterized by relatively short transfer arcs and modest $\Delta v$ requirements. Table~\ref{TABENG} lists S/C and engine data (namely dry mass $m_{dry}$, thrust magnitude $T$, specific impulse $I_{sp}$ and input power $P_{in}$). The trajectory design framework itself is not tied to a specific mission architecture, target, or propulsion system. All mission design parameters and constraints, including departure and arrival windows, stay times, and propulsion parameters, can be modified in a straightforward way, without altering the underlying structure of the method.\\ 
\indent The impulsive maneuvers of round-trip solutions $S_1$ and $S_2$ portrayed in Fig.~\ref{FIG:2SOLI} are replaced by low-thrust arcs using the method described in Sect.~\ref{ILT}, with maximum thrust step duration $\Delta t_{{max}} = 1$ hour and tolerance on the final mass ratio mismatch $\epsilon = 10^{-4}$. The two thrust segments are joined through a coast arc once the residual total impulse of the Lambert arc drops below $\Delta v_{th} = 1 \; \nicefrac{\mathrm{m}}{\mathrm{s}}$. The resulting low-thrust round-trip transfers are shown in Fig.~\ref{FIG:2SOLLT}, where a prescribed number of equally spaced arrows indicates the thrust direction along each thrust arc. In the present case, the rendezvous and takeoff masses are assumed to be equal (i.e., $m_2 = m_3$), neglecting the mass of any prospective asteroid samples. Tables~\ref{TABPERFIab} and \ref{TABPERFLTab} present the event dates and performance metrics for solutions $S_1$ and $S_2$ in the impulsive and low-thrust case, respectively, under the assumption $m_5 = m_{\mathrm{dry}}$, with redundant data omitted from Table~\ref{TABPERFLTab}. 

\subsection{Discussion}\label{DC}

\indent The modular structure of the technique, and in particular its ability to target specific PLOs around the Sun-Earth collinear libration points, naturally supports a variety of advanced mission architectures. A key capability is the systematic selection of the Sun-Earth gate ($L_1$ or $L_2$) and transfer type (MT or TO) for both Earth escape and return, something that is impossible to explicitly enforce with classical PC approaches. Once a S/C reaches a PLO, it can remain there for an arbitrary stay time with modest station-keeping effort, awaiting a new departure window either to revisit the same NEO or to rendezvous with a different one. This feature makes PLOs or similar libration point orbits ideal parking locations for future orbiting stations near the Sun-Earth and Earth-Moon libration points, in line with the current Gateway concept. In this context, the trajectory database produced by the strategy can be interpreted as a library of round-trip opportunities associated with a set of mission design parameters and constraints, each indexed by the departure and arrival PLO, Sun-Earth gate, transfer type, and the associated performance metrics. Complex mission architectures can be assembled by linking compatible round-trip transfers according to specific objectives. In more involved scenarios, the PLO index becomes particularly important, since a S/C returning to a given PLO must depart from the same PLO on its next leg. However, transitioning between different PLOs can be achieved with limited propellant consumption by exploiting the natural dynamics of the Sun-Earth system \cite{HidayJohnston1996TimeFree,Gomez1998HaloTransfer}.\\
\indent This perspective naturally supports a network of S/C operating from orbiting stations located in periodic orbits around the Sun–Earth $L_1$ and $L_2$ libration points. In this framework, different S/C could be assigned complementary mission roles, ranging from reconnaissance and scientific characterization to resource extraction or planetary-defense activities. 

\begin{table}[t!]
\centering
{\renewcommand{\arraystretch}{1.2}
\footnotesize
\caption{Minimum $\Delta v$ impulsive solutions to 1991 VG and 2014 DJ80 found in this work and in NHATS \cite{CNEOSNHATS}.} 
\label{TABPERFCOMP}
\begin{tabular}{l|cc|cc}

Object & \multicolumn{2}{c|}{1991 VG (index 2)} & \multicolumn{2}{c}{2014 DJ80 (index 20)} \\
\hline
Source & Ref.~\cite{CNEOSNHATS} & This work & Ref.~\cite{CNEOSNHATS} & This work \\
\hline
$C_{3}$ $\left[ \frac{\mathrm{km^2}}{\mathrm{s^2}} \right]$ & 2.985 & 0.239 & 2.057 & 0.018 \\
$\Delta v$ $\left[ \frac{\mathrm{m}}{\mathrm{s}}\right]$ & 1112 & 1273 & 2030 & 1349 \\
$v_{\infty}$ $\left[ \frac{\mathrm{m}}{\mathrm{s}} \right]$ & 1589 & 390 & 2896 & 428 \\
$\Delta t$ [day] & 386 & 1236 & 386 & 1122 \\
$\Delta t_{S}$ [day] & 32 & 251 & 112 & 166 \\
Launch date & 2039-May-01 & 2036-Apr-01 & 2041-Sep-03 & 2039-Oct-04 \\
\makecell[l]{Solution class} & \makecell[c]{PC} & \makecell[c]{$L_1$ MT–$L_2$ MT} & \makecell[c]{PC} & \makecell[c]{$L_2$ MT–$L_1$ MT} \\
\hline
\end{tabular}
}
\end{table} 

\indent Several large-scale surveys in the literature investigate round-trip transfers to NEOs. NHATS \cite{CNEOSNHATS} is selected as the benchmark for the solutions computed in this work. Note that NHATS trajectories correspond to high-energy mission scenarios (better suited for crewed missions to NEOs), whereas the low-energy round-trip transfers investigated in this study are more appropriate for unmanned probes. NHATS uses the PC method for the trajectory of the S/C with full-precision JPL ephemerides for Earth and NEOs, and automatically processes newly discovered objects and those with updated orbits. Table \ref{TABPERFCOMP} collects the minimum $\Delta v$ impulsive trajectories to asteroids 1991 VG and 2014 DJ80 obtained in this study and in NHATS. In addition to rendezvous and takeoff impulses, NHATS solutions include the maneuvers required to depart from a 400-km altitude geocentric circular parking orbit and to control atmospheric entry speed at Earth return. NHATS takes into account all four maneuvers to determine the minimum $\Delta v$ transfer. Rendezvous and takeoff impulses are joined in a single metric ($\Delta v$ in Table \ref{TABPERFCOMP}) and compared directly, while the characteristic launch energy $C_3$ and the return hyperbolic excess speed $v_{\infty}$\footnote{$\;$ For NHATS solutions, $v_{\infty}$ is calculated using the vis-viva equation with the re-entry speed extracted from \cite{CNEOSNHATS} and distance equal to the 6500-km atmospheric entry radius \cite{barbee2010comprehensive}.} are used for departure and arrival, respectively.\\ 
\indent Table \ref{TABPERFCOMP} intends to report the main properties of the two distinct NEO mission scenarios, while emphasizing that the new strategy proposed in this work is substantially different and particularly suitable for unmanned missions with extended stay time. With respect to NHATS trajectories, low-energy solutions are characterized by similar $\Delta v$ but significantly lower $C_3$ and $v_{\infty}$, which comes at the cost of increased durations due to the long time spent inside the TC along TOs or MTs. Because of the same reason, the departure epoch for the mission scenarios found in this work must be anticipated with respect to the respective launch dates found in NHATS. Comparison of Fig. \ref{FIG:SOLDV} with pork chop plots provided by \cite{CNEOSNHATS} for 1991 VG and 2014 DJ80 suggests that the proposed methodology expands departure and arrival windows, providing more launch and return opportunities compared to conventional PC designs.\\
\indent Unlike NHATS trajectories, which are fast and therefore better suited for chemical propulsion, a notable advantage of the low-thrust solutions found in this work consists in their limited propellant expenditure. Using the same 600-kg dry mass and a specific impulse of 400 $\mathrm{s}$ for chemical propulsion, the launch masses for the impulsive solutions in Table \ref{TABPERFCOMP} can be calculated using the Tsiolkowsky rocket equation. The launch mass in NHATS is 797 kg for 1991 VG and 1007 kg for 2014 DJ80, while in this work it is 830 kg for 1991 VG and 846 kg for 2014 DJ80. After replacing the impulses with low-thrust arcs, the launch mass becomes 663 kg for 1991 VG and 657 kg for 2014 DJ80, both lower than their NHATS counterparts.

\subsection{Round-trip mission to Apophis}\label{RTMA} 

\indent Many interesting and scientifically valuable NEOs on more eccentric or inclined orbits are filtered out by the target selection criteria adopted in Sect.~\ref{TS}, and therefore fall outside the analyzed population. This does not reflect a limitation of the proposed strategy, but rather an intentional preliminary reduction of the available object pool to facilitate the identification of round-trip transfers associated with limited total impulse, i.e., reduced propellant consumption. Indeed, the technique described in this work can be applied to any object, provided that at least one geometrical intersection between the ellipse of the S/C and that of the NEO exists. To illustrate how the methodology behaves when applied to targets outside the selected population, a representative round-trip mission to Apophis is examined. Apophis is a well-known NEO whose orbit lies just beyond the region of space accessed by MTs and TOs ($a = 0.922$ au, $e = 0.191$, and $i = 3.34$ degree).\\ 
\indent Because Apophis travels a more eccentric orbit than those of the NEOs in the target population, substantially higher rendezvous and takeoff impulses are expected. Accordingly, the upper limits on these maneuvers were increased to accommodate values in the multi-kilometer-per-second range. The larger transfer $\Delta v$ demands for a propulsion system capable of providing more thrust and better performance overall. In particular, the 600-kg dry mass is maintained, but the engine parameters are changed to those of the Advanced Electric Propulsion System developed by NASA and Aerojet Rocketdyne in its 9-kW power setting \cite{shastry202512}. Table~\ref{TABENGAPO} lists the new engine data. If longer transfer times were allowed (for instance, by admitting multi-revolution heliocentric paths), lower thrust settings would be sufficient to provide the total impulse required by the mission. 

\begin{table}
\centering
\caption{Engine data used for the mission to Apophis.}
\begin{tblr}{Q[c,m]|Q[c,m]|Q[c,m]}
$T$ & $I_{sp}$ & $P_{in}$ \\
\hline 
444 mN & 2605 s & 9 kW \\
\end{tblr}
\label{TABENGAPO}
\end{table}

\begin{table}
\centering
\caption{Minimum $\Delta v$ impulsive solutions to Apophis found in this work and in NHATS \cite{CNEOSNHATS}.} 
\begin{adjustbox}{max width=\textwidth}
{\renewcommand{\arraystretch}{1.2}
\footnotesize
\label{TABPERFCOMP_AST}
\begin{tabular}{l|cc}
Source & Ref.~\cite{CNEOSNHATS} & This work \\
\hline
$C_{3}$ $\left[ \frac{\mathrm{km^2}}{\mathrm{s^2}} \right]$ & 29.361 & 0.066 \\
$\Delta v$ $\left[ \frac{\mathrm{m}}{\mathrm{s}}\right]$        & 1594 & 4934 \\
$v_{\infty}$ $\left[ \frac{\mathrm{m}}{\mathrm{s}} \right]$     & 3909 & 7 \\
$\Delta t$ [day]                         & 354 & 1287 \\
$\Delta t_{S}$ [day]                     & 8 & 328 \\
Launch date                             & 2029-Apr-11 & 2043-Jan-07 \\
\makecell[l]{Solution class}           & \makecell[c]{PC} & \makecell[c]{$L_2$ MT-$L_2$ MT} \\
\hline
\end{tabular}
}
\end{adjustbox}
\end{table} 

\begin{figure}
    \centering
    \includegraphics[width=0.45\textwidth]{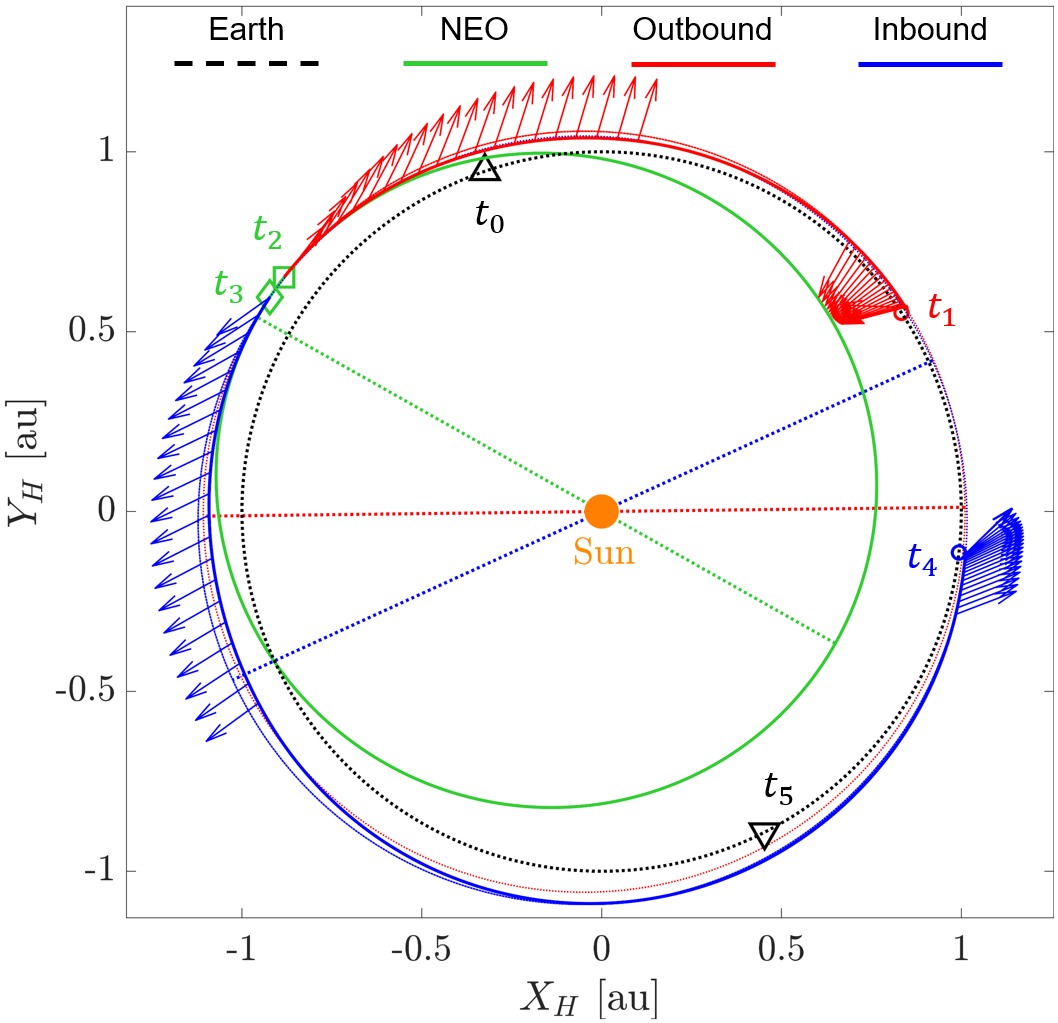}
    \vspace*{-3.5mm}
    \caption{Low-thrust round-trip transfer to Apophis with arrows showing thrust direction.} 
    \label{FIG:2SOLLTApophis}
\end{figure} 

\indent The performance of the minimum $\Delta v$ impulsive solution obtained in this work is compared with that computed by NHATS to the same object in Table~\ref{TABPERFCOMP_AST}. It must be emphasized that the NHATS transfer was retrieved using unconstrained settings and is drawn from a large window spanning 2020 to 2045. Moreover, according to the NHATS formulation, the impulse required to reduce $v_{\infty}$ to secure the re-entry does not contribute to the total impulse of the mission. As a direct consequence, the NHATS solution in Table~\ref{TABPERFCOMP_AST} is characterized by large values of $v_{\infty}$, close to the upper limit set by NHATS. In practice, the NHATS trajectory achieves significantly lower rendezvous and takeoff impulses by transferring most of the required energy change into $C_3$ and $v_{\infty}$. In contrast, the solution obtained in this work is inherently characterized by low departure and arrival energies, and therefore the rendezvous and takeoff impulses represent the major contributions to the overall velocity change.\\
\indent Figure \ref{FIG:2SOLLTApophis} portrays the low-thrust round-trip transfer to Apophis. The rendezvous and takeoff masses are assumed to be equal (i.e., $m_2 = m_3$) and $m_5 = m_{\mathrm{dry}}$. Using the same 600-kg dry mass and a specific impulse of 400 $\mathrm{s}$ for chemical propulsion, the launch masses for the impulsive solutions in Table \ref{TABPERFCOMP_AST} can be calculated using the Tsiolkowsky rocket equation. The launch mass is 901 kg in NHATS and 2110 kg in this work. Despite the significantly larger $\Delta v$ with respect to NHATS, the use of low thrust reduces propellant consumption, lowering the launch mass to 847 kg, approximately 50 kg lighter than NHATS. Most notably, in this case the advantage in terms of $C_3$ and $v_{\infty}$ is outstanding (see Table \ref{TABPERFCOMP_AST}). The lower launch mass, escape energy, and hyperbolic excess speed at re-entry demonstrate the strength of the proposed methodology compared to conventional PC approaches, even when applied to targets beyond its current intended scope.\\

\section{Concluding remarks}\label{CR}

\indent This study presents a systematic and effective approach for the design of low-energy round-trip trajectories to Near-Earth Objects (NEOs) based on the joint use of the Sun–Earth–S/C circular restricted three-body problem (CR3BP) and the Sun–spacecraft two-body problem (2BP). Building upon earlier work that addressed NEO rendezvous trajectories, this contribution extends the strategy to round-trip missions by independently computing outbound (Earth–NEO) and inbound (NEO–Earth) transfers, and subsequently patching together segments reaching and departing from the same object. Mission constraints and design parameters are included in straightforward way, and can be adjusted without altering the underlying structure, which is desirable in the early phases of mission analysis.
The methodology described in this work offers the
systematic generation of a large variety of round-trip solutions within a low-energy framework.
Moreover, impulsive transfers are converted into low-thrust trajectories, with apparent advantages in propellant consumption, while timing of NEO rendezvous and encounters with Earth are preserved.

More specifically, the proposed framework leverages natural pathways for Earth departure and return, significantly reducing launch and return costs. The autonomous nature of the CR3BP eliminates the need to predefine launch, NEO rendezvous/departure, and return epochs, which emerge naturally when mapping three-body trajectories to their heliocentric 2BP counterparts.\\
\indent The application of the method to a selected population of 80 dynamically accessible NEOs produced more than two million distinct round-trip trajectories. The resulting solution space spans a wide range of stay times, mission durations, and total impulses, enabling flexible trade-offs according to specific mission requirements. When compared with NASA's NHATS database, the proposed trajectories represent a distinct and alternative option, characterized by wider departure and arrival windows and significantly lower characteristic launch energies and Earth-return hyperbolic excess speeds, while maintaining rendezvous and departure impulses of similar magnitude.\\
\indent Unmanned missions can further benefit from the long transfer times naturally associated with low-energy trajectories through the use of low thrust. Once impulsive round-trip trajectories are computed, they are seamlessly transformed into low-thrust transfers by replacing discrete impulsive maneuvers with continuous thrust arcs while preserving event dates. To do this, a dedicated algorithm is proposed, developed, and successfully applied. The resulting low-thrust solutions require modest power and mass budgets, remaining well within the capabilities of existing power and propulsion technologies, and supported by data from recent deep-space missions. The combination of low thrust and low-energy three-body trajectories reduces launch and return costs, propellant consumption, and widens departure and arrival windows, improving mission flexibility.\\ 
\indent Beyond the construction of single-object round-trip trajectories, the modularity of the proposed framework supports the design of advanced mission architectures that exploit the dynamical structures of the Sun–Earth system. This can include repeated visits to a group of objects, as well as missions aimed at NEO hazard mitigation. 
Moreover, Sun-Earth libration orbits can be used as staging locations, with also the prospect of recurrent missions to selected, especially interesting targets.
It is worth remarking that the technique described in this paper can be  adapted to high-energy trajectories, suitable for shorter (crewed) missions. Future work will focus on extending the method to three-dimensional trajectories, in the context of a high-fidelity dynamical framework.

\begingroup
\section*{Acknowledgments}
\indent This research was funded by Khalifa University of Science and Technology through the Competitive Internal Research Award under Project ID: KU-INT-CIRA-2021-65/8474000413. This research was supported by Polar Research Center (PRC), Khalifa University of Science and Technology (KU-PRC). E. Fantino further acknowledges the support of grant PID2021-123968NB-I00 (Spanish Ministry of Science and Innovation).
\endgroup


\end{document}